\documentclass{emulateapj}

\begin{document}

\title{MHD waves in two-dimensional prominences embedded in coronal arcades}

\author{J. Terradas, R. Soler, A. J. D\'\i az, R. Oliver, J. L. Ballester}
\affil{Departament de F\'\i sica, Universitat de les Illes Balears, E-07122 Palma de Mallorca, Spain}
\email{jaume.terradas@uib.es}

\begin{abstract}

Solar prominence models used so far in the analysis of MHD waves in such
structures are quite elementary. In this work, we calculate numerically
magnetohydrostatic models in two-dimensional configurations under the presence
of gravity. Our interest is in models that connect the magnetic field to the
photosphere and include an overlying arcade. The method used here is based on a
relaxation process and requires solving the time-dependent nonlinear ideal MHD
equations. Once a prominence model is obtained, we investigate the properties of
MHD waves superimposed on the structure. We concentrate on motions purely
two-dimensional neglecting propagation in the ignorable direction. We
demonstrate how by using different numerical tools  we can determine the period
of oscillation of stable waves. We find that vertical oscillations, linked to
fast MHD waves, are always stable and have periods in the 4-10 min range.
Longitudinal oscillations, related to slow magnetoacoustic-gravity waves, have
longer periods in the range of 28-40 min. These longitudinal oscillations are
strongly influenced by the gravity force and become unstable for short magnetic
arcades. \end{abstract}


\maketitle

\section{Introduction}

It is well known since the 1960s that solar prominences and filaments show
oscillations \citep[see][]{ramsmith66,hyder66}. Many of these oscillatory
motions have been classified as large-amplitude oscillations. A clear example of
these kind of oscillations are the global motions found in winking filaments.
These global oscillations are normally induced by nearby sub-flares or jets, EIT
waves and Moreton waves. The reader is referred to \citet{tripatetal09} for a
review about observations of large amplitude oscillations in prominences.

The MHD eigenmodes of oscillation of prominences in simple geometries as
Cartesian slabs and cylindrical magnetic tubes, have been studied in the past by
many authors \citep[see the reviews
of][]{oliball02,oliver2009,mackayetal10,arreguietal12}. These studies have
focused on small amplitude oscillations using the linearized version of the
MHD equations. Only recently,  \citet{bloklandkeppens11} have attempted to
understand localized MHD oscillations in more realistic configurations. The aim of our
work is to study global oscillations using improved prominence models which are
numerically constructed using a relaxation method. At this stage we are not
interested on the internal fine structure of prominences and  our focus is
mainly on the global behavior and the possible link with winking filaments. 

An issue that arises in complex magnetic topologies under the presence of gravity
is stability. Here we conduct an MHD stability study that enables us to
understand the stable/unstable nature of quiescent prominences. The stability
analysis is not simple. A method to study the stability properties of the new
equilibria is to compute the full ideal or resistive MHD spectrum by solving the
linearized MHD equations. Another alternative is to consider the time dependent
problem by solving the nonlinear or the linearized MHD equations. This last
approach, successfully used in the past in linear stability analysis of, for
example, coronal arcades \citep[see][]{anetal89}, is adopted in the present work.
There are other possibilities such as as the variational or energy method
\citep[see][]{bersteinetal1958} which is based on the minimization of the second
order change in the potential energy of the system when plasma elements are
displaced from their equilibrium position. Another alternative is to use
magneto-frictional methods \citep[see][]{yangetal86} based on the assumption that
field lines move through a stationary medium. This method has been successfully
used in the determination of  non-linear force-free coronal field in response to
the evolution of the photospheric magnetic field \citep[see for
example][]{mackayvan06,mackayvan09}.

It is worth to mention that \citet{galindotrejo87} performed a detailed
numerical stability analysis of two-dimensional prominence models based on known
analytical MHS solutions at that time (Kippenhahn-Schl\"uter, Dungey, Menzel and
Lerche \& Low models). However, the connection of the prominence magnetic field
with the photospheric magnetic field was essentially missing in his analysis. In
this work, we properly address this point, which turns to be very relevant
regarding the stability of prominences. Later, \citet{debruynehood93}
demonstrated that the model of \citet{low81} is unstable to localized
disturbances and that the \citet{hoodanzer90} model is only stable for
sufficiently low prominences. 

In the literature many prominence models have been proposed \citep[see the
review of][]{mackayetal10}. A popular model is the magnetic flux rope
configuration. Using this configuration \cite{lowzhang04} found analytical
solutions using a polytropic model in a circular cylinder whose weight is
supported by an external magnetic field. Later, \citet{petrieetal07}
demonstrated how to numerically calculate magnetohydrostatic equilibria with
properties close to realistic prominences, namely a cool dense prominence
surrounded by a cavity within a flux rope in a coronal environment.
\citet{bloklandkeppens11a} solved an extended Grad-Shafranov equation, and using
a finite element-based code were able to obtain numerical equilibria.
\citet{bloklandkeppens11} used these equilibria to analyze the continuous
spectrum of modes of the structure. These authors focused on the modes of the
core of the prominence rather than on the modes of the global structure since
the 2D flux rope considered in their studies does not curve down and meet the
photosphere (a 3D model is required to fulfill this condition in flux rope
models).

In this work we avoid geometries with detached magnetic field lines, i.e., we 
study configurations with all field lines tied to the lower boundary. Detached
models are usually considered in the study of 2D twisted flux rope prominence
models. We prefer to concentrate on configurations that connect magnetic field
lines to the photosphere. This can be also achieved considering twisted flux
ropes in 3D and using, for example, toroidal geometries. Nevertheless, we think
that it is more convenient to start with the investigation of the 2D problem
rather than with the full 3D problem, which is also more complicated from the
technical point of view. Additionally, we want to address the role of magnetic
dips in the structure and also the dynamics of prominences. For this reason, we
chose a topology which includes magnetic dips that are able to provide suitable
conditions to support the cool plasma against gravity  \citep[see for
example][]{demoprist93,aulanier02,lopezaristeetal06,mackayetal10}.

The purpose of this paper is first to construct prominence models by computing 
magnetohydrostatic (MHS) solutions of the MHD equations. We seek for  prominence
models that are bounded in the 2D plane and have a cool core respect to the
external coronal environment, meaning that the structure is non-isothermal. We
focus our attention on models that describe both the prominence and the
surrounding coronal environment under the presence of gravity. Here the
prominence model is constructed using a relaxation process instead of the direct
solution of the Grad-Shafranov equation. Mass is injected on an initial
background equilibrium and the system is allowed to evolve towards a new
equilibrium. We are not aiming to study the formation process itself. Instead,
we are interested on the final MHS solution. The second goal of this work is to
study the properties of MHD waves in the numerically generated prominence
models. The time-dependent problem is solved numerically. An initial
perturbation is introduced in the system and it excites different kinds of
oscillations which might be stable or unstable.

\section{Initial configuration and setup}

\subsection{Background equilibrium}\label{backequil}

The primary equilibrium is an isothermal stratified atmosphere permeated by a
force-free magnetic field. Using a Cartesian coordinate system, with the
$z-$coordinate pointing in the vertical direction, the density profile is 
\begin{eqnarray}\label{dens0}
\rho=\rho_{\rm 0}\,e^{-z/\Lambda},
\end{eqnarray}
\noindent where ${\Lambda}=c_{\rm s0}^2/\gamma g$ is the density scale height and
$\rho_{\rm 0}$ is the coronal density value at the reference level $z=0$
representing the photosphere or base of the corona. The sound speed, defined as $\sqrt{\gamma
p_0/\rho_0}$, 
takes a value of $c_{\rm s0}=166 \,\rm km\, s^{-1}$ for a coronal temperature of
$10^6\,\rm K$. The gravity acceleration on the solar surface is $g=0.274 \,\rm
km\, s^{-2}$ and for a monoatomic gas $\gamma=5/3$. Hereafter, we choose a
spatial reference length of $H=10^{4}\,\rm km$ (the typical length of
prominences) meaning that the density scale height is $\Lambda\approx 6
H$. 

The initial potential force-free magnetic field considered in this work is based
on superposition of arcade solutions. The arcade configuration  has the following magnetic field
components 
\begin{eqnarray}\label{bxp} B_x(x,z)&=&B_0 \cos{k x}\, e^{-k\, z},\\ 
B_z(x,z)&=&-B_0\sin{k
x}\, e^{-k\, z},\label{bzp} 
\end{eqnarray}
\noindent where $B_0$ is the magnetic field strength at the reference level. The parameter $k$ is related to
the lateral extension of the arcade ($\pi/(2k)$) and is also a measure of the vertical
magnetic scale height. The $B_y$ component is zero in the present work.

The magnetic field lines in the configuration given by
Eqs.~(\ref{bxp})-(\ref{bzp}) do not have any dips because the magnetic structure
is bipolar. Since we are interested in a configuration with dips for the reasons
explained in the Introduction, we select a particular superposition of two
magnetic arcades that mimics a quadrupolar configuration 
\begin{eqnarray}\label{bx} B_x(x,z)&=&B_1\cos{k_1 x}\, e^{-k_1\,
z}-B_2\cos{k_2 x}\, e^{-k_2\, z},\\
B_z(x,z)&=&-B_1\sin{k_1 x}\, e^{-k_1\, z}+B_2\sin{k_2 x}\, e^{-k_2\, z}.\label{bz} \end{eqnarray}
The individual arcade solutions are quoted with the sub-indices 1 and 2.  The
width of the full structure is $2L$ and we select the following wavenumbers
 $k_1=\pi/(2L)$ and $k_2=3\pi/(2L)$. The strength of the magnetic field at $z=0$ 
of each arcade is $B_1$ ($>0$) and $B_2$ ($>0$). From the superposition of the
two configurations it is easy to show that at $z=0$ the total magnetic field has
a maximum value at $x=\pm L$ of $B_{\rm max}=B_1+B_2$.

\begin{figure}[!h] \center{\includegraphics[width=8cm]{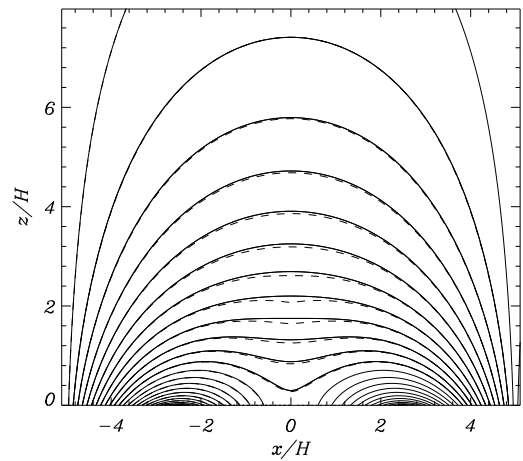}} 
\caption{\small Magnetic field lines based on a quadrupolar magnetic field. In
this plot $B_2=B_1$, $k_1=\pi/(2L)$, $k_2=3\pi/(2L)$ have been used in
Eqs.~(\ref{bx})-(\ref{bz}). Here $L=5 H$, being $H$ the reference length
($10^4\,\rm km$). Solid curves correspond to the case without a dense
prominence, while dashed curves are the new equilibrium structure after the dense
material, representing a prominence, has been injected.}\label{magneticfield}
\end{figure}

\noindent An example of the magnetic configuration for the case $B_2=B_1$ is
shown in Fig.~\ref{magneticfield}.  At the center of the magnetic configuration
($x=0$) there is an $X-$point where the magnetic field in the $xz-$plane is
zero. In the example of Fig.~\ref{magneticfield} the location of this point is
at $z=0$. In general, it can be shown that the height of the $X-$point is 
\begin{eqnarray}\label{zx} z_{\rm X}=\frac{1}{k_2-k_1}\ln \frac{B_2}{B_1}.
\end{eqnarray}
\noindent In this work we always impose that the $X-$ point is at the
photospheric level meaning that $B_2=B_1$.

\begin{figure}[!ht] \center{\includegraphics[width=7cm]{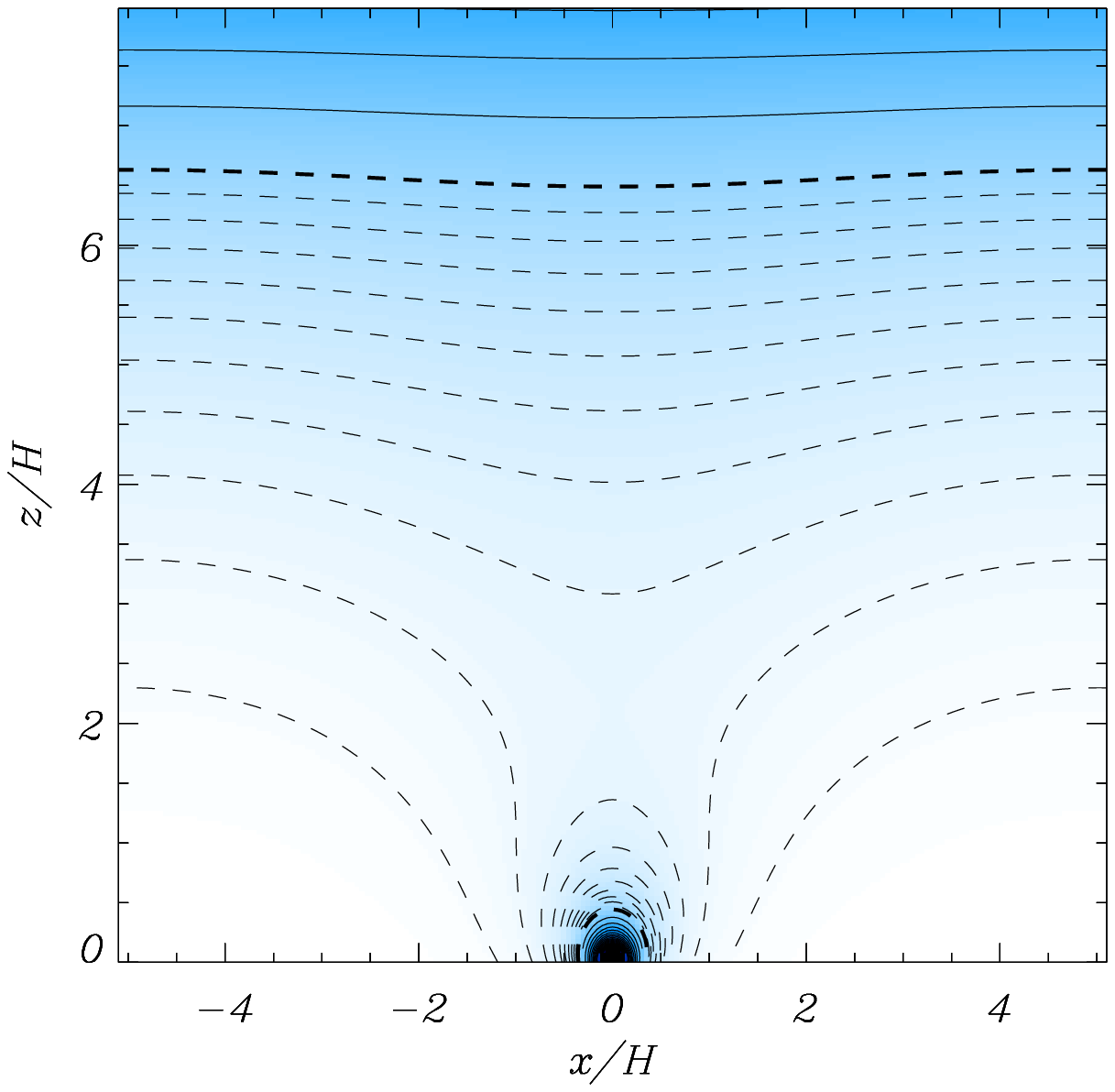}}
\center{\includegraphics[width=7cm]{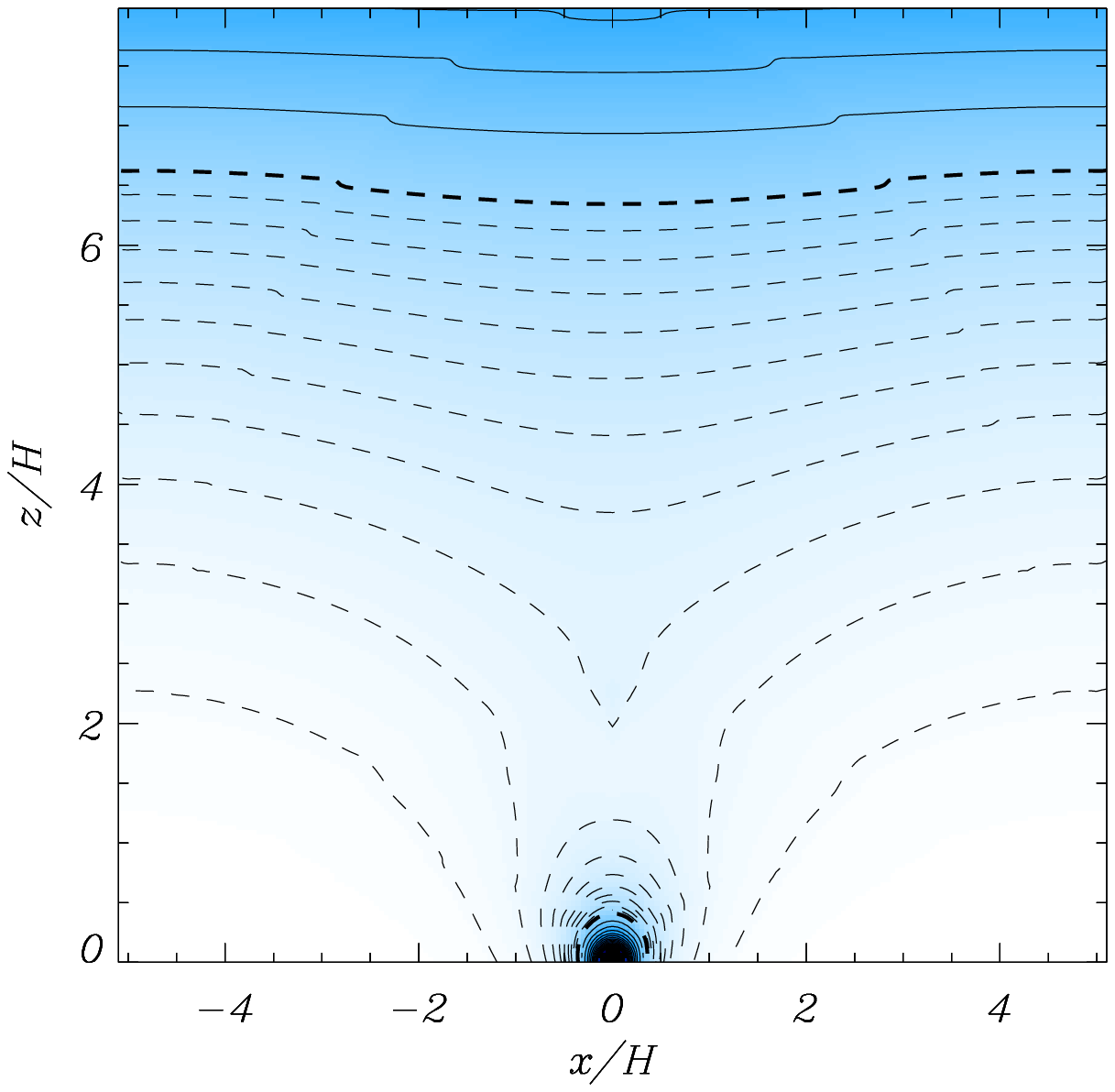}}
\center{\includegraphics[width=7cm]{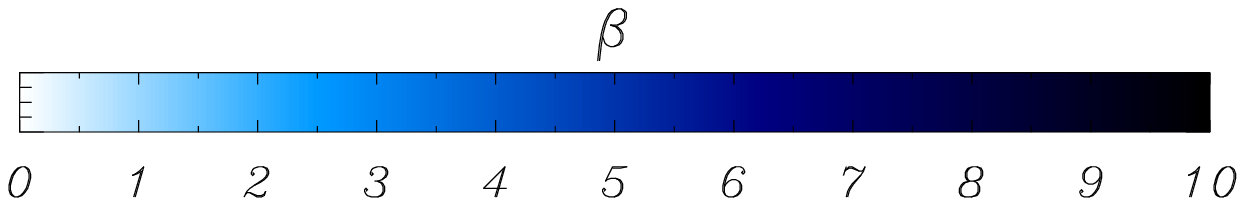}} \caption{\small
Plasma-$\beta$ corresponding to the case without prominence (top panel)
associated to the solid curves in Fig.~\ref{magneticfield} and to the situation
after the dense material has been injected (bottom panel), corresponding to the
dashed curves in Fig.~\ref{magneticfield}. The injection point is around $x=0$
and $z=2H$. Continuous curves represent values of $\beta$ above 1, while dashed
curves correspond to $\beta$ lower than 1. The thick dashed curve corresponds to
$\beta=1$. $\beta$ tends to infinity at $x=z=0$ in this 2D potential configuration
since there is no magnetic shear.}\label{plasmabeta} \end{figure}

The Alfv\'en speed, $v_{\rm A}=B/\sqrt{\mu_0 \rho}$, is easily computed given the
magnetic field configuration (Eqs.~(\ref{bx})-(\ref{bz})) and the density
profile (Eq.~(\ref{dens0})). For normalization purposes the Alfv\'en speed is
normalized to the maximum value at $z=0$ and $x=\pm L$, and it is referred here
as $v_{\rm A0}$.  Another useful magnitude is the plasma-$\beta$  defined as
$\beta=2 c_{\rm s}^2/\gamma v_{\rm A}^2$ and represented in the top panel of
Fig.~\ref{plasmabeta} for the magnetic structure of Fig.~\ref{magneticfield}.
Since in this case the field is strictly zero at the $X-$point, the
plasma-$\beta$ tends to infinity at $x=z=0$. From Fig.~\ref{plasmabeta} we
see that the plasma-$\beta$ is less than one in most of the spatial domain. The
prominence body will be located around $x=0$ and $z=2\,H$.

\subsection{Mass deposition}\label{inject}

To obtain a model resembling a real prominence a dense and cool plasma is
required. In the present work we are not concerned about the actual physical
process that provides mass to the prominence during its formation. Instead, here
we are interested in finding an equilibrium configuration for the prominence and
we do not care about the actual process that drives the formation
\citep[see][for recent results about the formation
process]{xiaetal11,xiaetal12,lunaetal12a}.

A simple way to generate the body of the prominence is to add mass at a given
location in the preexisting magnetic configuration. We model the mass injection by
artificially adding a source term in the continuity equation.
This term has the following form
\begin{eqnarray}\label{source}
S={\hat \alpha} \sin\left(\pi \frac{t}{t_m}\right)
e^{-\left(x/w_x\right)^2-\left(\left(z-z_0\right)/w_z\right)^2}.
\end{eqnarray}
The parameter $\hat \alpha$ represents the rate of mass injection and $t_m$ is
the total injection time. For $t>t_m$ the source term is set to zero. The
parameters $w_x$ and $w_z$ represent the characteristic spatial size of the
source in the $x$ and $z-$directions respectively. The central point of the injection is located  at $x=0$ and
$z=z_0$.

From Eq.~(\ref{source}) it is straight forward to calculate the total mass
injected in the system. Note that the configuration studied in this work is two-dimensional and thus
unbounded in the $y-$direction. For this reason it is more convenient to calculate
the total mass of the prominence per unit length, $M/L_y$. After the injection phase we
have that 
\begin{eqnarray}\label{totalmass} 
M/L_y&=&\frac{2\,\hat \alpha\, t_m} {\pi }\int_{\rm -\infty}^{\infty} \int_{\rm -\infty}^{\infty}
e^{-\left(x/w_x\right)^2-\left(\left(z-z_0\right)/w_z\right)^2}
dx\, dz \nonumber\\
&\approx&2\,\hat \alpha\, t_m\, w_x\, w_z.
\end{eqnarray}
\noindent From observations it is possible to roughly estimate the total mass
of a real prominence. The length along the main axis is usually a known
parameter. For a real prominence it can be shown that $M/L_y\sim 5\times 10^5\,
\rm kg\, km^{-1}$. To get this value we have used a typical prominence density
of  $5\times 10^{-2}\,\rm kg\, km^{-3}$, while for the spatial dimensions we chose
that the width is $10^{3}\,\rm km$, the height is $10^{4}\,\rm km$, and its
length ($L_y$) is around $10^{4}\,\rm km$. In our computations we use the length
$H=10^{4}\,\rm km$, already introduced before, as the reference length. Another important parameter in our
model is the size of the arcade ($2L$) that provides the magnetic support. We
assume, based on observations, that the typical length of the magnetic field lines
are of the order of $10^{5}\,\rm km$. For a given value of $M/L_y$ and size of
the prominence ($w_x$ and $w_z$) the  product of $\hat \alpha$ and $t_m$ is
determined using Eq.~(\ref{totalmass}). From the practical point of view we fix
the parameter  $t_m$ and calculate  $\hat \alpha$ for a given prominence mass and
size. The reference time-scale in this
work is defined as $\tau_{\rm A}=H/c_{\rm s0}$ which for $H=10^4\,\rm km$ and
$c_{\rm s0}=166\, \rm km\,s^{-1}$ is around $1\,\rm min$.

\section{Numerically generated MHS prominence models}\label{newequil}

\subsection{Numerical tools}

Using the initial background model, explained in Section \ref{backequil}, the
injection of  mass is performed employing the specific profile described in
Section \ref{inject}. The nonlinear ideal MHD equations are advanced in time
numerically using the code MoLMHD \citep[see][for details about the numerical
method]{bonaetal09,terrolietal08}. The source term included in the continuity equation
provides the mass required to generate a prominence model. It is important to
mention that although the full nonlinear equations are solved we only evolve in
time perturbations on the background magnetic field \citep[see][]{powelletal99}.
It turns out that this numerical technique is crucial to obtain new
magnetohydrostatic models for low plasma-$\beta$ problems.

Boundary conditions are treated using a decomposition in characteristic variables
at the edge of the computational domain. The different fields are recalculated at
the boundary by imposing conditions on the incoming fields. Line-tying
conditions, applied at $z=0$, meaning that incoming fields are set to be equal to
outgoing fields. For the rest of the domain, flow-through conditions are applied
imposing that incoming fields are set to zero. Different sizes of the domain in
the $x$ and $z-$directions have been considered but we have found that the
results do not significantly depend on the extension of the computational domain.
The use of conditions on the characteristic fields ensured minimal reflections
from the lateral and top edges of the domain, and perfect reflection at the
bottom of the computational box. Moreover, in order to obtain solutions that are
close to the static stationary state for some specific cases we have used the
decomposition in characteristic variables to eliminate perturbations in the
system. In particular, we have found that imposing flow-through conditions only
on slow MHD modes and on the entropy mode helps the system to relax faster to the
stationary state. In this case, line-tying conditions have been applied to
fast and Alfv\'en MHD waves at $z=0$.

A linearized version of the code has been also used in the study of linear MHD
waves. We have found that in this case it is not necessary to  use decomposition
in characteristic fields at the boundaries since simple reflection conditions at
the photosphere and flow-through conditions at the lateral and top edges perform
well.

The simulations have been carried out using grids of typically $400\times 400$
points in the $x$ and $z-$directions. We have found that the results converge if
the resolution is raised above the previous numbers. The code MoLMHD runs in
parallel using MPI and the computational facilities of the Solar Physics Group at
UIB have been used to perform the simulations. The simulation time for a run with
32 processors is around 2 hours.

\subsection{Properties of the generated MHS equilibria}

\begin{figure}[!ht] \center{\includegraphics[width=8cm]{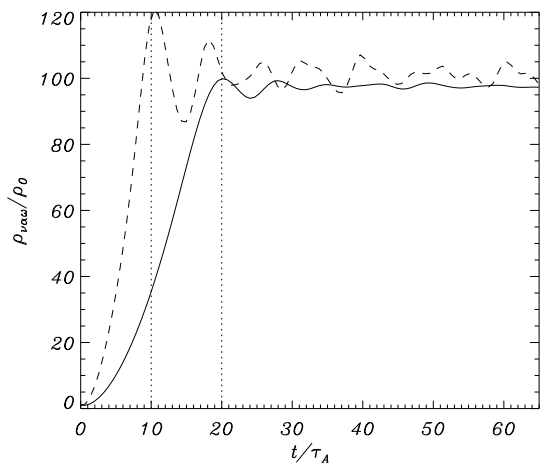}} 
\caption{\small Maximum density as a function of time. This maximum is achieved
around the point $x=0$ and $z=2 H$. The dashed curve corresponds to the case
$t_m=10\, \tau_{\rm A}$ and $\hat \alpha=10$, while for the continuous curve
$t_m=20\, \tau_{\rm A}$ and $\hat \alpha=5$.  The vertical dotted lines
denote the time when the mass injection is switched off and are associated to the
dashed curve ($t_m=10\, \tau_{\rm A}$) and continuous curves ($t_m=20\, \tau_{\rm
A}$).  The total mass injected is the same for the two simulations. In all the
simulations $w_x=0.2/\sqrt{2}\,H$ and $w_z=0.5/\sqrt{2}\,H$ and $z_0=2
H$.}\label{graph3} \end{figure}

We start by analyzing the results of the simulations, and we concentrate first on
the maximum density in the system localized at the center of the prominence core.
For the present case the location of this maximum is not far from the center of
the mass deposition (located at $x=0$ and $z=z_0$) but for configurations with
higher $\beta$ the maximum is located at lower heights. The results are plotted
in Fig.~\ref{graph3} (see continuous curve). The maximum density grows smoothly
with time due to the source term in the continuity equation. This source term is
set to zero for $t>t_m$ ($t_m=20\, \tau_{\rm A}$ in the present case). At this
stage $\rho_{\rm max}$ shows an almost constant value, suggesting that the system
has reached or it is very close to a new equilibrium state. The maximum
velocities, not shown here, are also rather small in the whole computational
domain,  typically of the order of $0.5\,\rm km\,s^{-1}$.

In the example shown in Fig.~\ref{graph3} we have not eliminated the reflection
of slow MHD waves at $z=0$. Therefore, oscillations in $\rho_{\rm max}$ after the
injection phase are mainly due to the excitation of small-amplitude slow MHD
waves. In Fig.~\ref{graph3} the results of a case with  $t_m'=t_m/2$ and $\hat
\alpha'=2 \, \hat \alpha$ are also displayed (dashed line). According to
Eq.~(\ref{totalmass}) the mass per unit length of the prominence is the same for
the two simulations. From Fig.~\ref{graph3} we see that now the oscillatory
behavior has a larger amplitude, indicating a stronger back reaction of the
system to the newly added mass. A fast mass injection produces a more complex
relaxation of the configuration, since MHD waves with higher amplitudes are
excited in the system. These oscillations are related to the eigenmodes of the
configuration and will be studied later. As we are not intending to model the
prominence formation itself we choose the parameter $\hat \alpha'$ in such a way
that we get to the  stationary state without large amplitude oscillations
involved in the transient phase.

\begin{figure}[!ht] \center{\includegraphics[width=7cm]{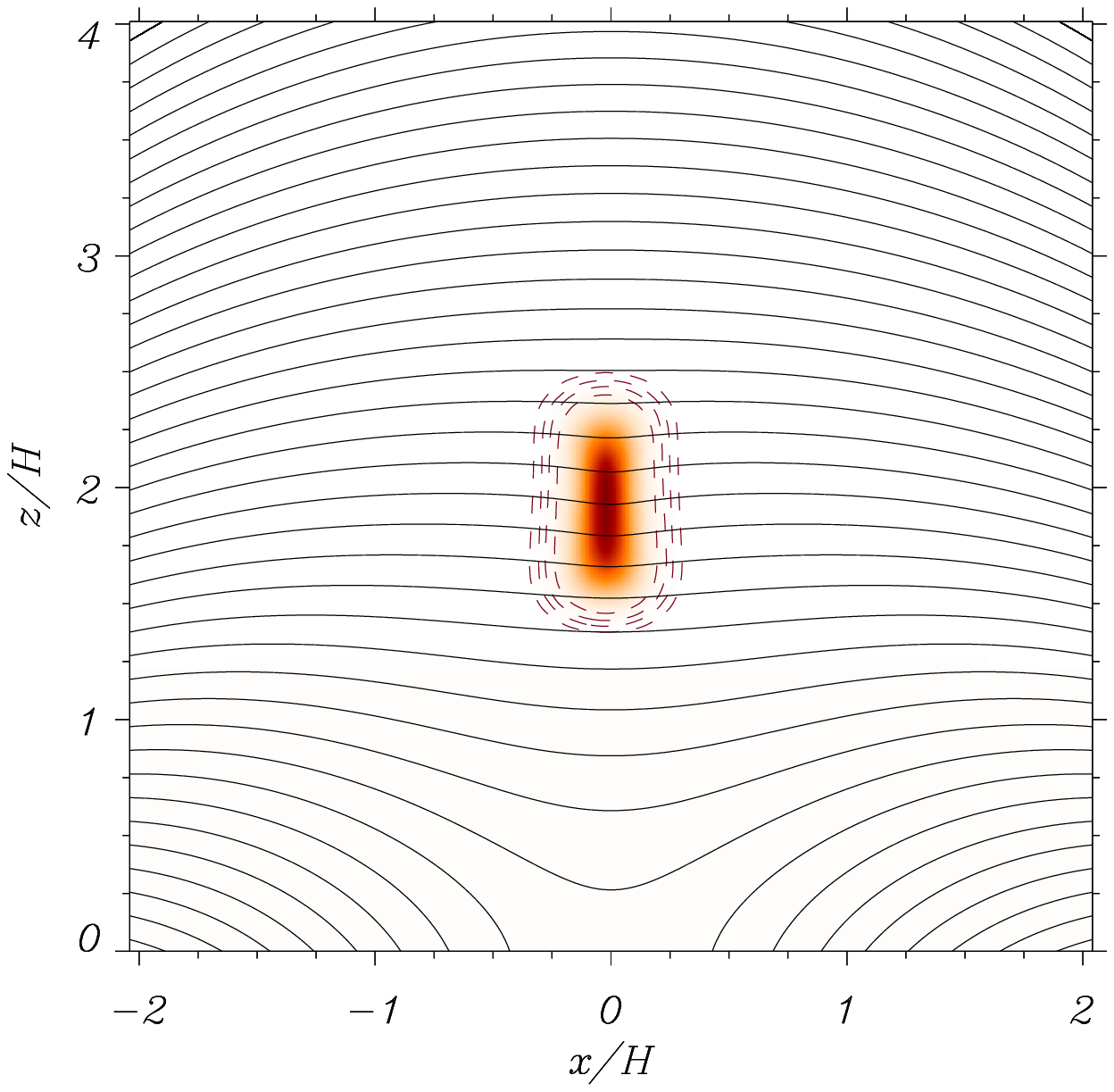}}
\center{\includegraphics[width=7cm]{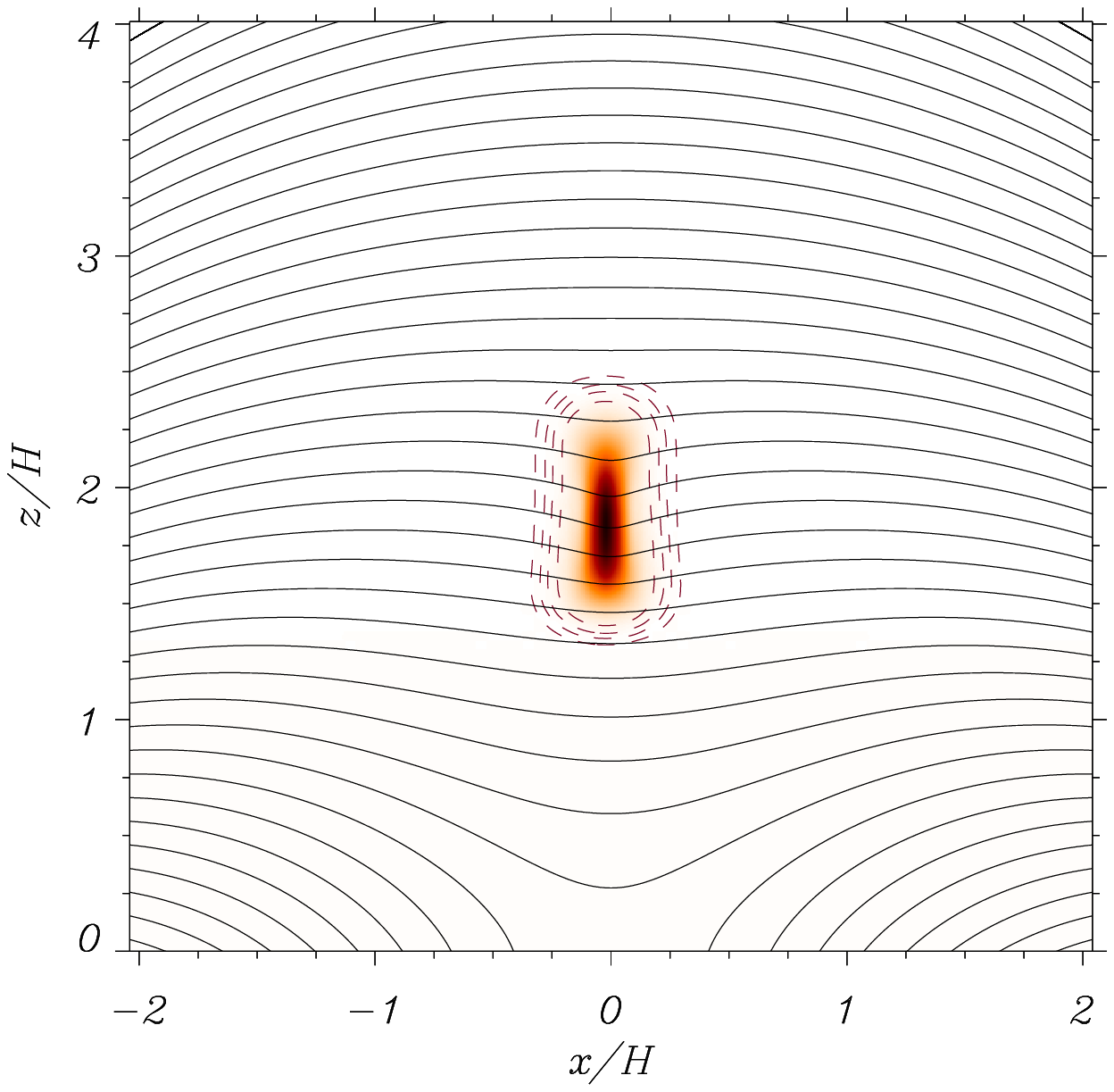}}
\center{\includegraphics[width=7cm]{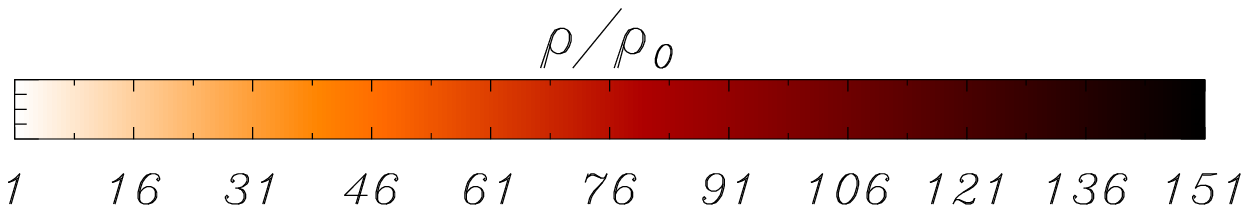}} \caption{\small Density
distribution, red color scale, representing a prominence in the stationary
state. Dashed curves are iso-contours of temperature while continuous
curves are the magnetic field curves. In the top panel the configuration
corresponds to the plasma-$\beta$ distribution shown in Fig.~\ref{plasmabeta}
(bottom panel), while in the bottom panel the plasma-$\beta$ is two times
larger.}\label{denstemp} \end{figure}

In Fig.~\ref{denstemp}, top panel, the two-dimensional density distribution is
plotted at a fixed time ($t=40\, \tau_{\rm A}$) in a reduced spatial domain of
$4H\times4H$. The injected mass has been redistributed in the structure and the
system is close to a new equilibrium, as Fig.~\ref{graph3} already indicates.
For this particular simulation the total mass per unit length is around
$5.4\times 10^{5}\,\rm kg\, km^{-1}$ and it is of the same order as the
reference value given in Section \ref{inject}. During the relaxation process the
gas pressure has changed and the Lorentz and gravitational forces have been
adapted to the newly added mass. For comparison with the initial magnetic field,
we overplot in Fig.~\ref{magneticfield} the new magnetic field configuration
(dashed curves). The deformation of magnetic field lines is clear. Due to the
presence of the heavy prominence the  magnetic field has been pushed down mostly
at the location of the enhanced density. The deformation of the magnetic
structure is not big, but enough to support, together with the pressure force,
the prominence. The dense material produces a depression in the Alfv\'en and
sound speeds. Note that the generated prominence is bounded in the $z-$direction
from below and from above, a feature that is not easily implemented in
theoretical models, with the exception of \citet{fiedlerhood92} who presented
numerical examples of 2D quiescent prominences with normal polarity modeled by a
cool isothermal slab of finite width and height.

\begin{figure}[!ht] \center{\includegraphics[width=7cm]{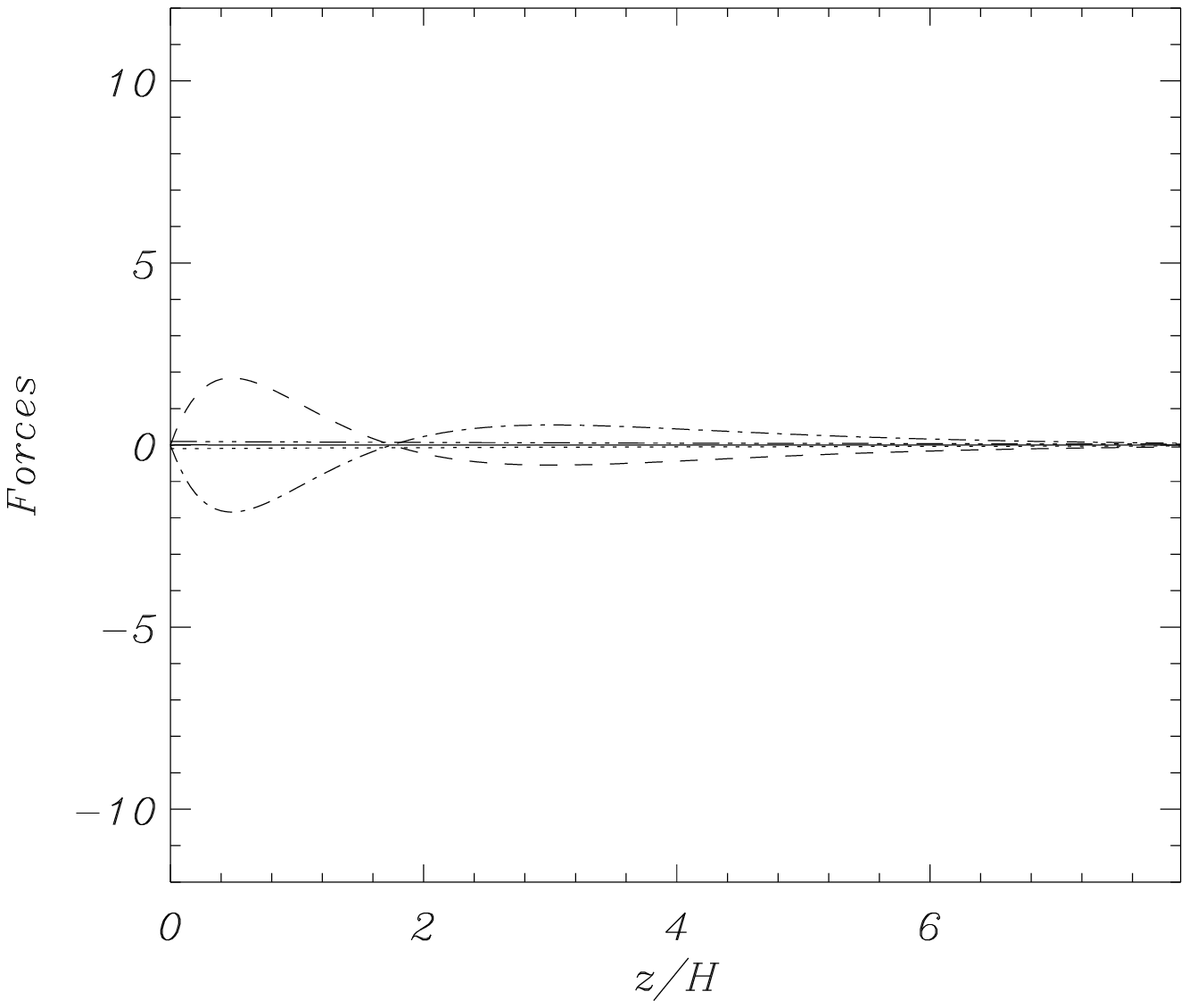}} 
\center{\includegraphics[width=7cm]{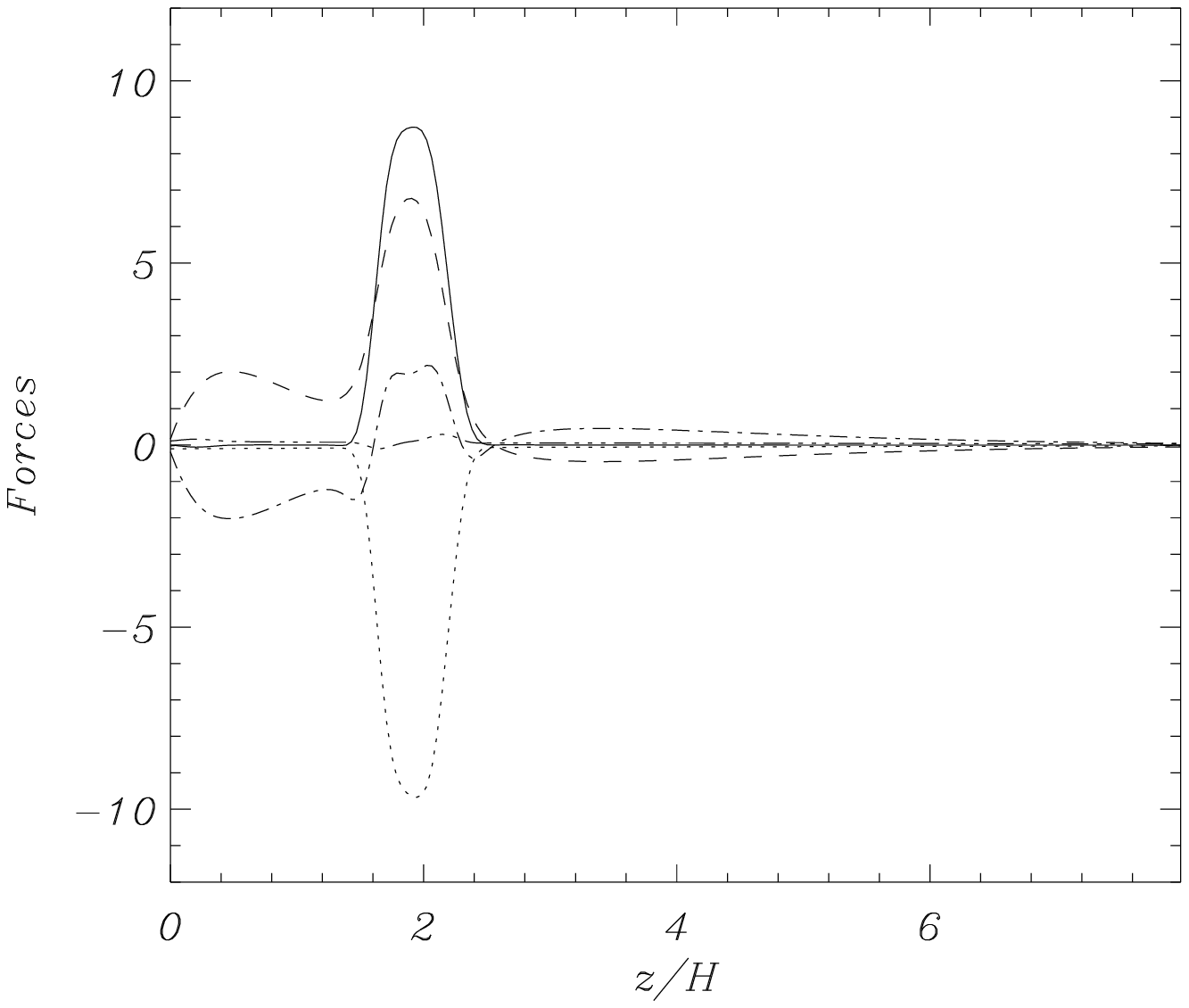}} \caption{\small Vertical
component of the forces  at $x=0$ as a function of height at the initial (top panel) and
close to the final state (bottom panel). The solid curve corresponds to the
total Lorentz force, dashed curves represent the magnetic tension and dot-dashed
curves correspond to the magnetic pressure gradient. The gravitational force is
plotted with dots and the pressure force with triple dot-dashed curves.}\label{forces} \end{figure}

The details about the distribution of the forces  in the vertical direction
when the system is close to an equilibrium are displayed in Fig.~\ref{forces}. In
this figure the different forces involved in the system are plotted as function
of height at the center of the prominence  ($x=0$). For comparison purposes we have also
included the initial configuration (top panel), characterized by a balance
between the pressure gradient and the gravity force. The initial magnetic
configuration is force-free and there is  a balance between the magnetic tension
and the gradient of the magnetic pressure. The final distribution of the forces
after the mass injection is displayed in the bottom panel of Fig.~\ref{forces}.
At the center of the prominence (around $z=2H$) the gravitational force is very
strong due to the presence of the dense material. This downward force is balanced
essentially by the upward Lorentz force, while the pressure gradient is still
small. The Lorentz force is mostly due to the tension term but the magnetic
pressure term has also a contribution to the total magnetic force (around 20$\%$
in Fig.~\ref{forces}). Just below the prominence body tension and magnetic
pressure forces are still larger than in the initial state, since the prominence
has pushed down the magnetic configuration increasing the depth of the dips of
the structure.

\begin{figure}[!ht] \center{\includegraphics[width=7cm]{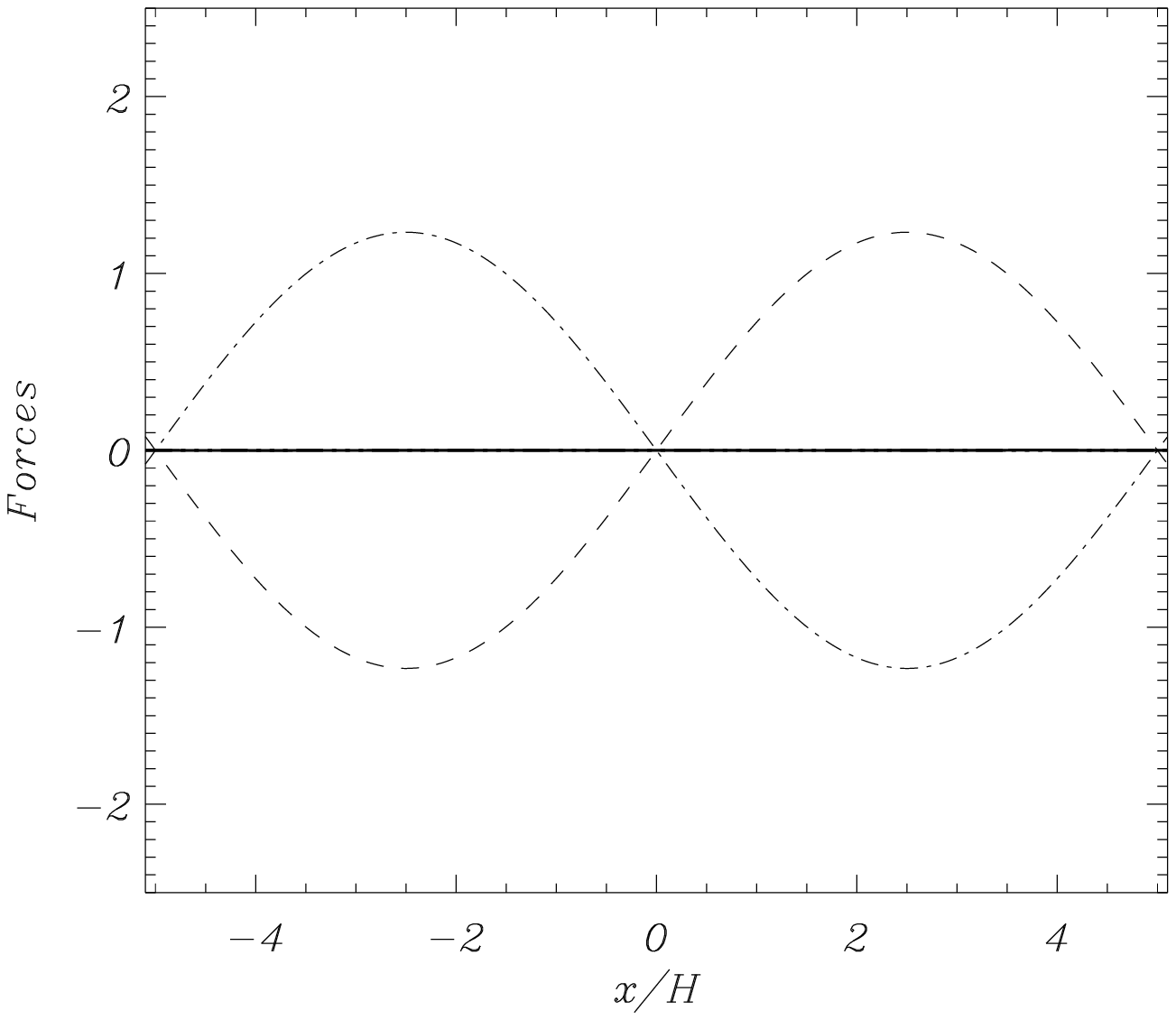}} 
\center{\includegraphics[width=7cm]{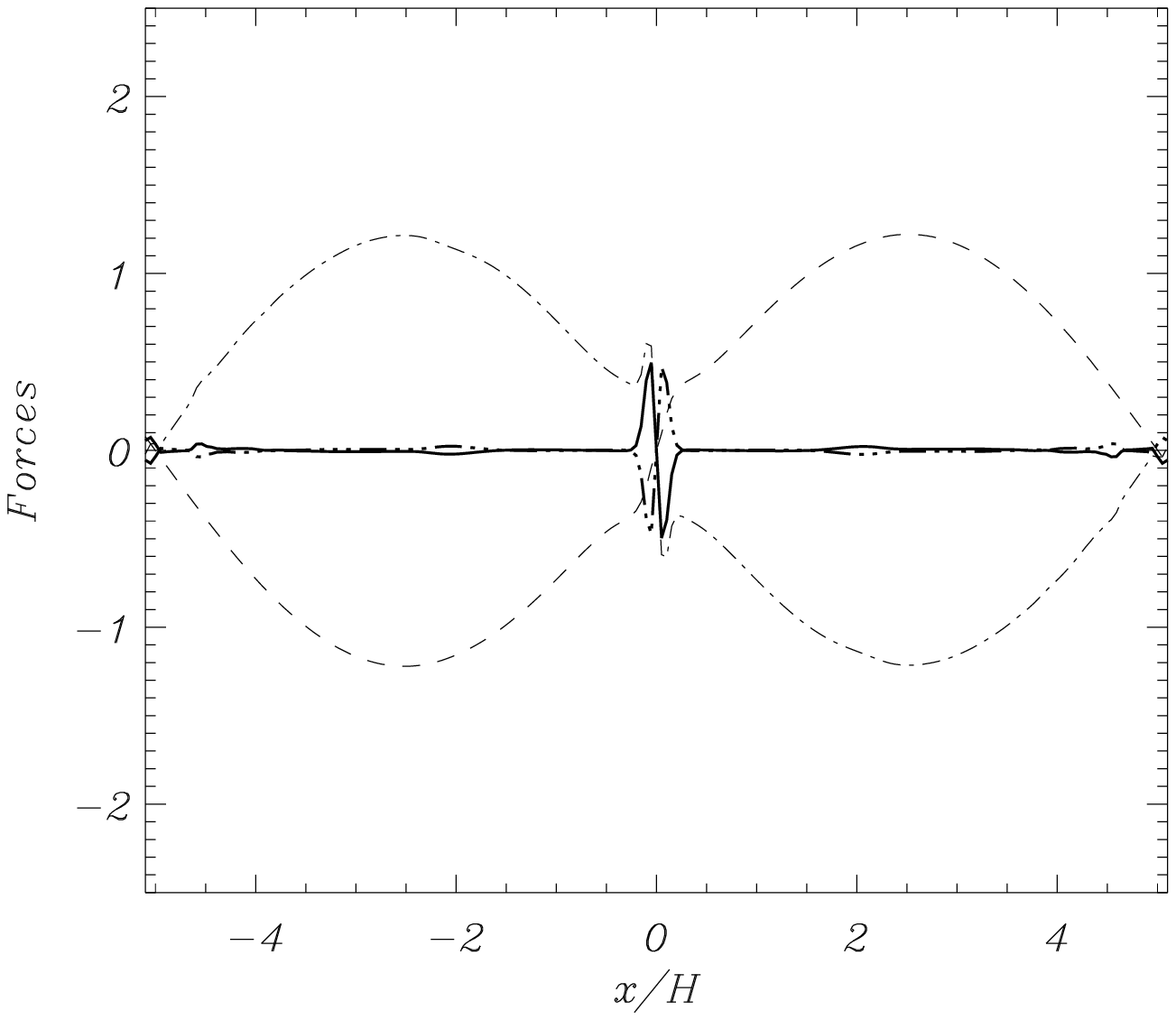}} \caption{ \small Horizontal
component of the forces at $z=2H$ as a function of the $x-$coordinate at the
initial (top panel) and close to the final state (bottom panel). The solid 
curve corresponds to the total Lorentz force, dashed curves represent the
magnetic tension and dot-dashed curves correspond to the magnetic pressure
gradient. The pressure force is plotted with triple dot-dashed
curves.}\label{forceshor} \end{figure}

The distribution of the forces in the horizontal direction is represented in
Fig.~\ref{forceshor}. Initially (top panel) the total Lorentz force is zero and
since gas pressure is constant in the horizontal direction the pressure gradient
is also zero. Once the mass has been injected the magnetic pressure gradient and
the gas pressure gradient change inside the prominence in such a way that they
balance each other.

\begin{figure}[!h] \center{\includegraphics[width=8cm]{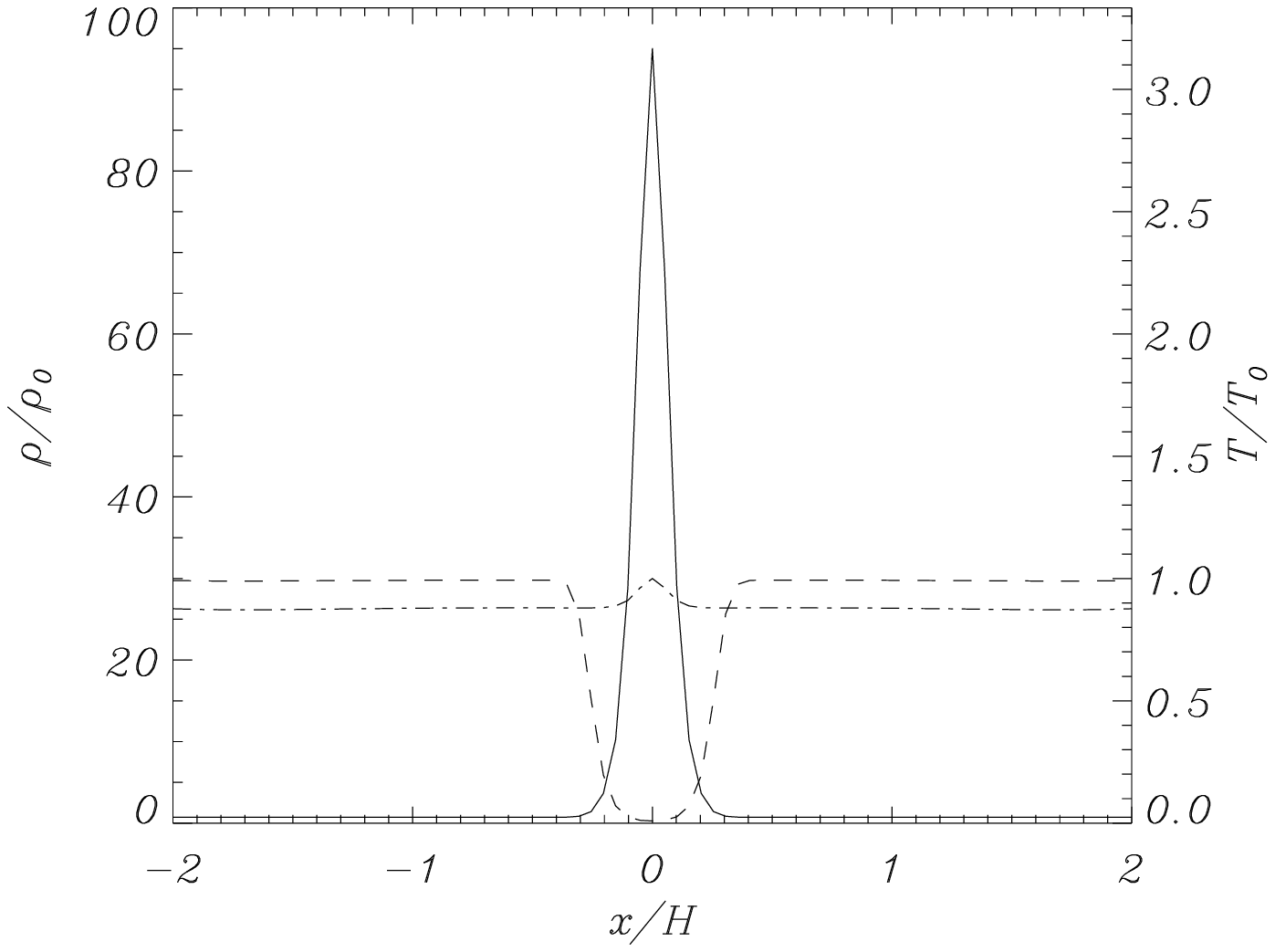}}
\center{\includegraphics[width=8cm]{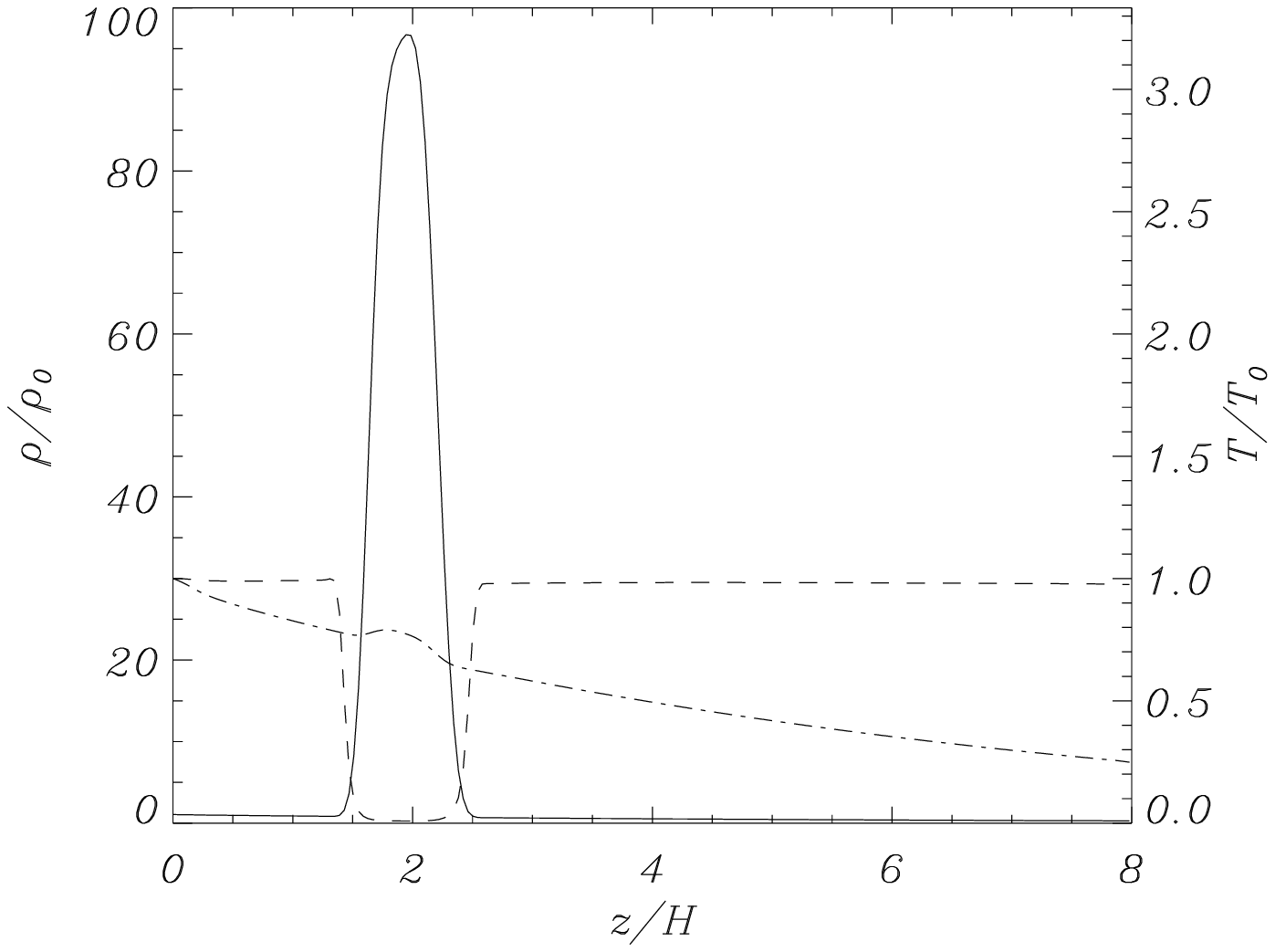}} \caption{\small Density
(continuous curve), temperature (dashed curve)  and gas pressure
(dot-dashed) distributions across the prominence body at $z=2H$ (top panel) and
along the center the prominence located at $x=0$ (bottom  panel). Density and
temperature are normalized to $\rho_0=5\times 10^{-4}\rm kg\, km^{-3}$, the
coronal density, and to the coronal temperature, $T_0=10^6\,\rm K$, respectively.
 Gas pressure is shown in arbitrary units in this plot. }\label{graph5}
\end{figure} 

The temperature at the center of the prominence after the mass injection  is
much lower than the initial temperature, while the gas pressure at the center of
the prominence does not change much with respect to the initial value, around
$14\%$ (see the temperature and density distribution in Fig.~\ref{graph5}).  This
can be understood from the behavior of the gas pressure, if pressure is
essentially constant at the center of the prominence then the dense part of the
prominence must be cooler than the light and hot coronal environment since
$\rho_P T_P \approx \rho_c T_c$. The temperature profile is not imposed in the
simulations and is self-adjusted during the injection phase. A cut of the
density, temperature,  and pressure across $z=2H$ is plotted in
Fig.~\ref{graph5} top panel. In this simulation the density reaches a maximum
value around 95 times the coronal density while the temperature should have,
according to the previous expression, a minimum value around 100 times lower than
the coronal temperature. The exact value for the temperature minimum at the core
of the prominence is $11,491\,\rm K$ which is of the same order of the
temperatures typically inferred from observations. In the bottom panel of
Fig.~\ref{graph5} the variation of density, temperature,  and pressure is
plotted as a function of height. The density enhancement representing the
prominence is superimposed to the exponentially decreasing profile of the
background model. The size of the prominence core and the corresponding PCTR is
determined by the form of the source term and the parameters $w_x$ and $w_z$.
Note that in the present model we ignore the effect of thermal conduction, which
may have an important impact on the shape of the PCTR.  Effects due to
radiative processes are also neglected in the present model.

\begin{figure}[!ht] \center{\includegraphics[width=7cm]{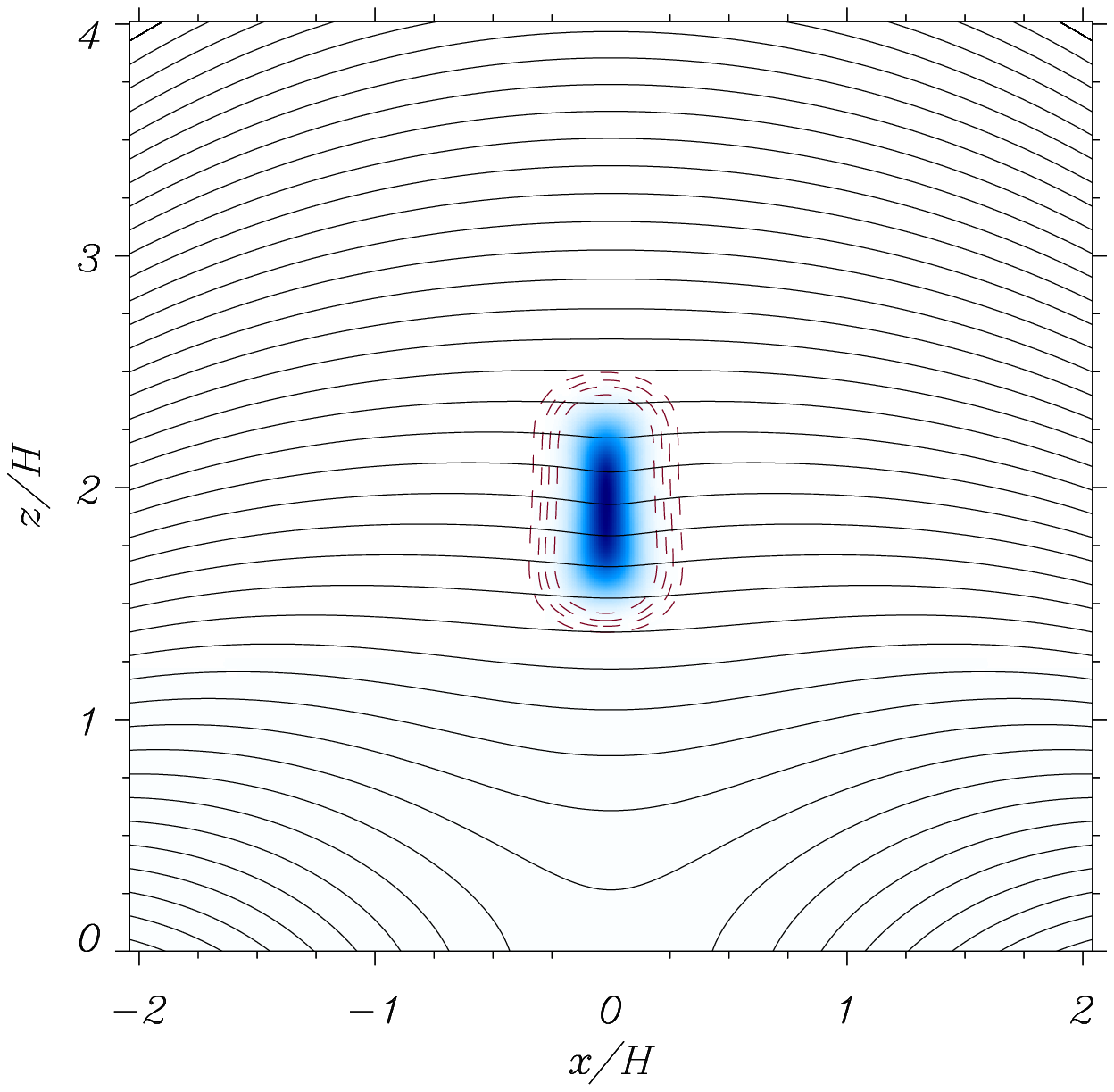}}
\center{\includegraphics[width=7cm]{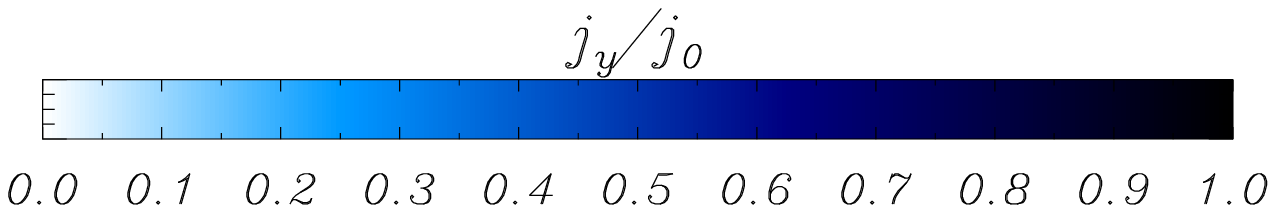}} \caption{\small
Current density in the $y-$direction, blue scale, calculated from the newly
generated model after the injection phase. The dashed curves represent
iso-contours of temperature. The continuous curves are the magnetic field
curves. This plot corresponds to the case shown in Fig.~\ref{denstemp} top
panel.}\label{graph6} \end{figure}

Although the initial magnetic field is potential the extra mass added to the
system makes the configuration non-potential. In Fig.~\ref{graph6} the current
density in the $y-$direction is displayed. This variable peaks at the core of
the prominence and reflects the fact that now the magnetic field  distribution
has been modified. It is interesting to note that the distribution of the
current is very similar to the density distribution (compare Fig.~\ref{graph6}
and and top panel of Fig.~\ref{denstemp}).

A change in the strength of the magnetic field produces a different final equilibrium. In Fig.~\ref{denstemp} bottom panel,
density, temperature and magnetic field distribution are plotted for a
plasma-$\beta$ two times larger than that of the case shown in the top panel.
Now the deformation of the field inside the prominence is more pronounced since
the Alfv\'en speed is $\sqrt{2}$ times smaller, meaning that tension and
magnetic pressure forces are weaker. The injected mass is able to strongly
modify the magnetic field configuration producing a magnetic structure more
curved at the core of the prominence. In fact, the center of the core of the
prominence is located at a lower height in comparison with the upper panel of
Fig.~\ref{denstemp}. The density at the core of the prominence is also larger.
The process of relaxation to this configuration lasts longer than that for the
lower $\beta$ case. We  have performed other experiments changing the
plasma-$\beta$ and found that for high values of this parameter very strong
shock waves are generated, indicating that the initial configuration is still
far from an equilibrium state. It is easier to obtain a new equilibrium if the
plasma-$\beta$ is low. This is in agreement with the the recent work of
\citet{hilliervan13} \citep[see also][]{anetal88,fiedlerhood92}.

\begin{figure}[!ht] \center{\includegraphics[width=8cm]{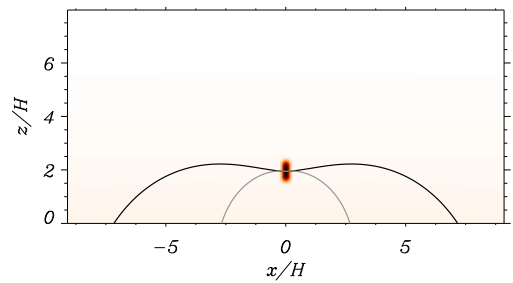}} 
\caption{\small Magnetic field lines at the center of the prominence for two
different prominence models. The black curve correspond to $L=9 H$, while for
the grey curve $L=3 H$. In this plot $v_{\rm A0}=10\,c_{\rm s0}$. The same color scale
for the density as in Fig.~\ref{denstemp} has been used.}\label{fieldlines_L}
\end{figure}

\begin{figure}[!ht] \center{\includegraphics[width=8cm]{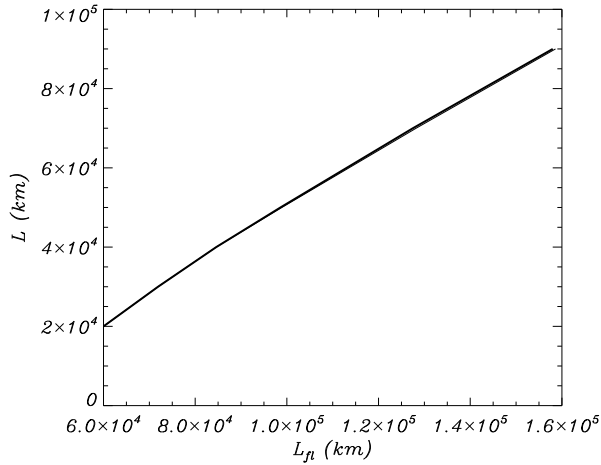}} 
\caption{\small Relation between the arcade half width and the total length of
the magnetic field line crossing the center of the prominence. In this plot 
$v_{\rm A0}=10\,c_{\rm s0}$.}\label{p_L-Llf} \end{figure}

Another relevant parameter in the models is the length of the arcade ($2 L$) in
which the prominence is embedded. In Fig.~\ref{fieldlines_L} the magnetic field
lines crossing the center of the body of the prominence are plotted for two
different values of $L$. In this example the large arcade is three times wider
than the short arcade. We have considered that $L$ is in the rage $2-9\times
10^4 \,\rm km$ in this work. For each value of $L$ in this range we have
calculated the total length of the magnetic filed line crossing the prominence
center, $L_{\rm fl}$. The relationship between $L$ and $L_{\rm fl}$ is plotted in
Fig.~\ref{p_L-Llf} and will be used later. 

\begin{figure}[!ht] \center{\includegraphics[width=8cm]{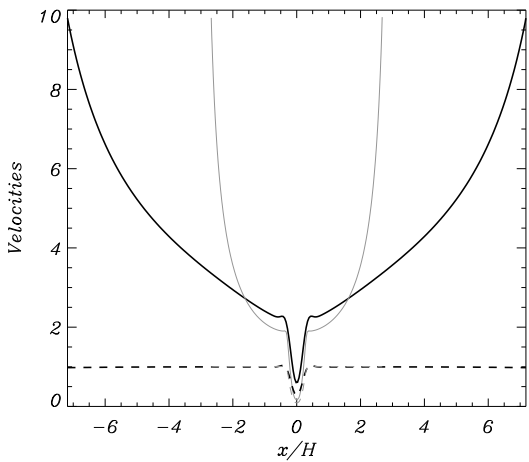}} 
\caption{\small Alfv\'en (continuous curve) and sound (dashed curve) speeds
(normalized to the background sound speed, $c_{\rm s0}$) as a function of the
$x-$coordinate along the field line crossing the center of the prominence and
displayed in Fig.~\ref{fieldlines_L}. The black curve correspond to $L=9 H$,
while for the grey curve $L=3 H$. In this plot $v_{\rm A0}=10\,c_{\rm s0}$.}\label{va_cs_L}
\end{figure}

Returning Fig.~\ref{fieldlines_L} notice that the dip in the magnetic field
lines is quite different in size for the two prominence models, and this can
have important implications regarding stability, as we will discuss later. These
two models have quite a different length of the field lines and also a different
variation of the equilibrium magnitudes. Alfv\'en and sound speeds are plotted
in  Fig.~\ref{va_cs_L} as a function of the $x-$coordinate for the two field
lines represented in Fig.~\ref{fieldlines_L}. In this plot we see that for the
narrow arcade the Alfv\'en velocity variation is stronger than for the wide
arcade, and that near the prominence body, characterized by the depression
around $x=0$, the Alfv\'en speed in the external medium is lower for the narrow
arcade. On the contrary, the profile of the sound speed is quite similar for the
two models. It is important to realize that the variation of the Alfv\'en speed
along the field lines is due to the change in both the modulus of the magnetic
field and density. In Fig.~\ref{va_cs_L} the Alfv\'en speed at the $z=0$ level
has been set to $v_{\rm A0}=10\,c_{\rm s0}$, i.e., $v_{\rm A0}=1666\,\rm km\,s^{-1}$. This
value determines a strength of the magnetic field at the base of the corona of
10 Gauss. Due to the variation with height of the quadrupolar magnetic
configuration the magnetic field strength decreases up to a value of around 2
Gauss at the core of the prominence. This is a limitation of our model because
to have higher values of the magnetic field at the prominence body requires a
significant increase of the Alfv\'en speed which is most likely unrealistic.
Here we have considered that at most $v_{\rm A0}=20\,c_{\rm s0}$, and this leads to a
value of 4 Gauss at the prominence center.

Finally, it is worth to mention that three different values for the total mass
of the prominence have been considered in this work. The values of $M/L_y$ are
$5.4$, $2.3$, and $1.3\times 10^{5}\,\rm kg\, km^{-3}$. This will allow us to
analyze the effect of the total mass on the periods of oscillation of the
different prominence models.

\section{Oscillations in the numerically generated models}

\begin{figure}[!ht]
\center{\includegraphics[width=8cm]{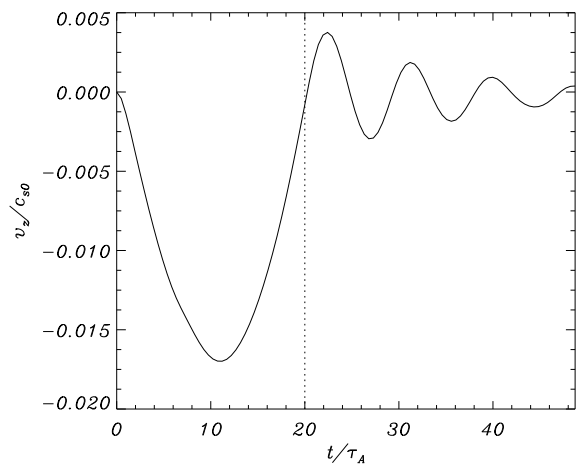}}  \caption{\small Vertical
velocity near the center of the prominence ($x=0$, $z=2H$) as a function of
time. The dotted vertical line denotes the time when mass injection is switched
off. For this simulation $v_{\rm A0}=10\,c_{\rm s0}$ and $L=5H$.}\label{graph7}
\end{figure}

The process of mass injection yields to the excitation of waves in the
configuration. Some of these waves have a strong leaky character and leave the
system quite quickly. However, there are other waves that are clearly associated
to oscillations of the density enhancement and are analyzed here using the
velocity field. A clear example of periodic oscillations is found in
Fig.~\ref{graph7}. In this plot the vertical component of the velocity is
plotted at a fixed point near the center of the prominence. We observe a
substantial variation of the amplitude which is associated to the mass injection
process, followed by short period waves. This periodic oscillation is damped
with time.  In fact, vertical oscillations take place before the new
equilibrium is reached so that the prominence is still moving downwards. In
Fig.~\ref{graph7} there is a small global negative velocity shift in the
vertical velocity after the injection phase, meaning that the whole structure is
not yet oscillating around the final equilibrium position, although it is quite
close to it since the drift tends to zero for long times. For this reason, we
think that to better understand waves in the configuration it is more convenient
to study oscillations once the prominence is in equilibrium. Hence, we let
evolve the system for long times till it relaxes to the stationary state. Then
we have two possibilities, either we introduce a perturbation in the system
using the full nonlinear MHD equations, or we simply focus on the linear problem
solving the linearized MHD equations around the final equilibrium. Here we adopt
the last approach.

\subsection{Vertical oscillations}\label{subsectvert}

\begin{figure}[!ht] \center{\includegraphics[width=7cm]{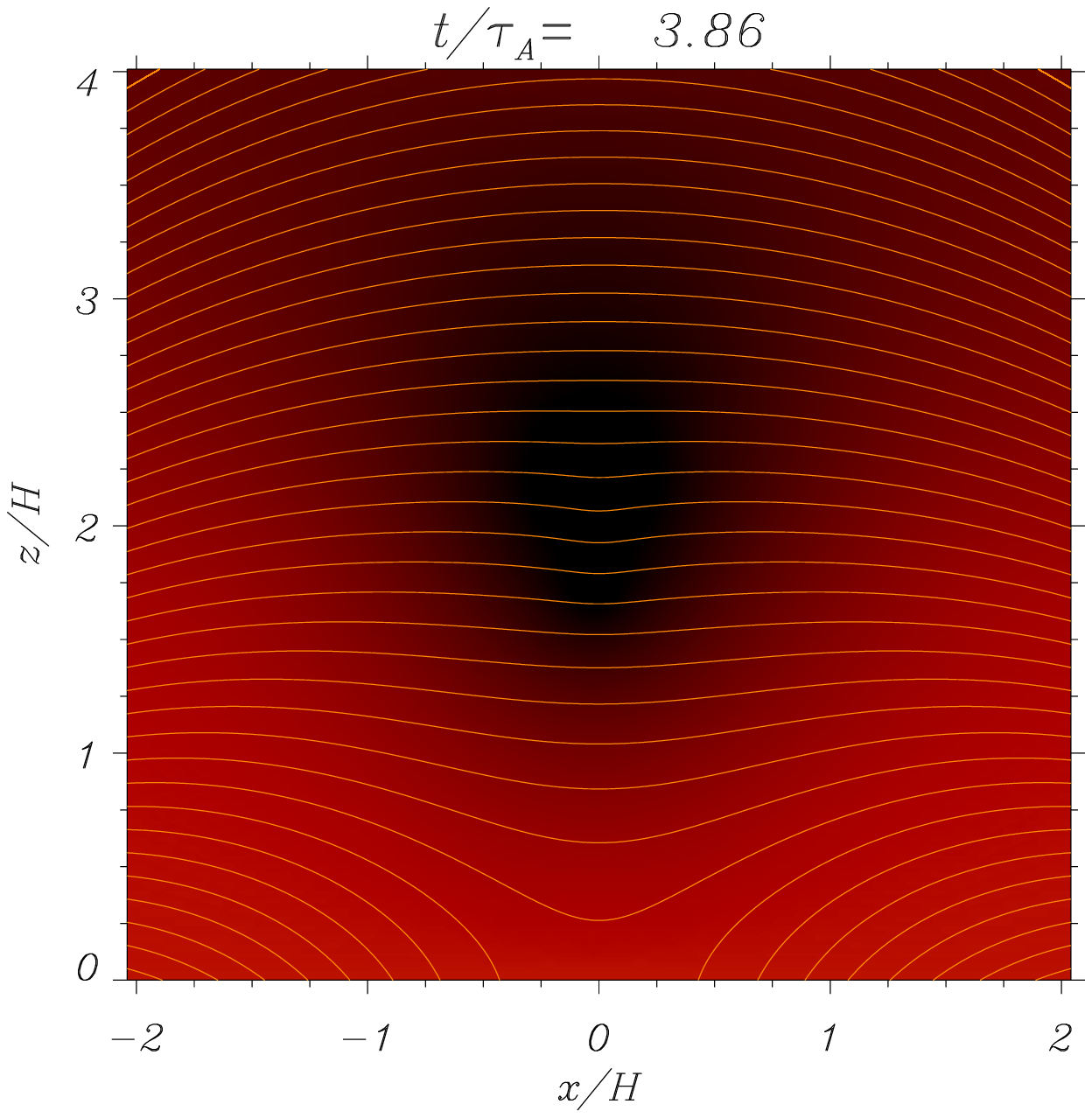}}
\center{\includegraphics[width=7cm]{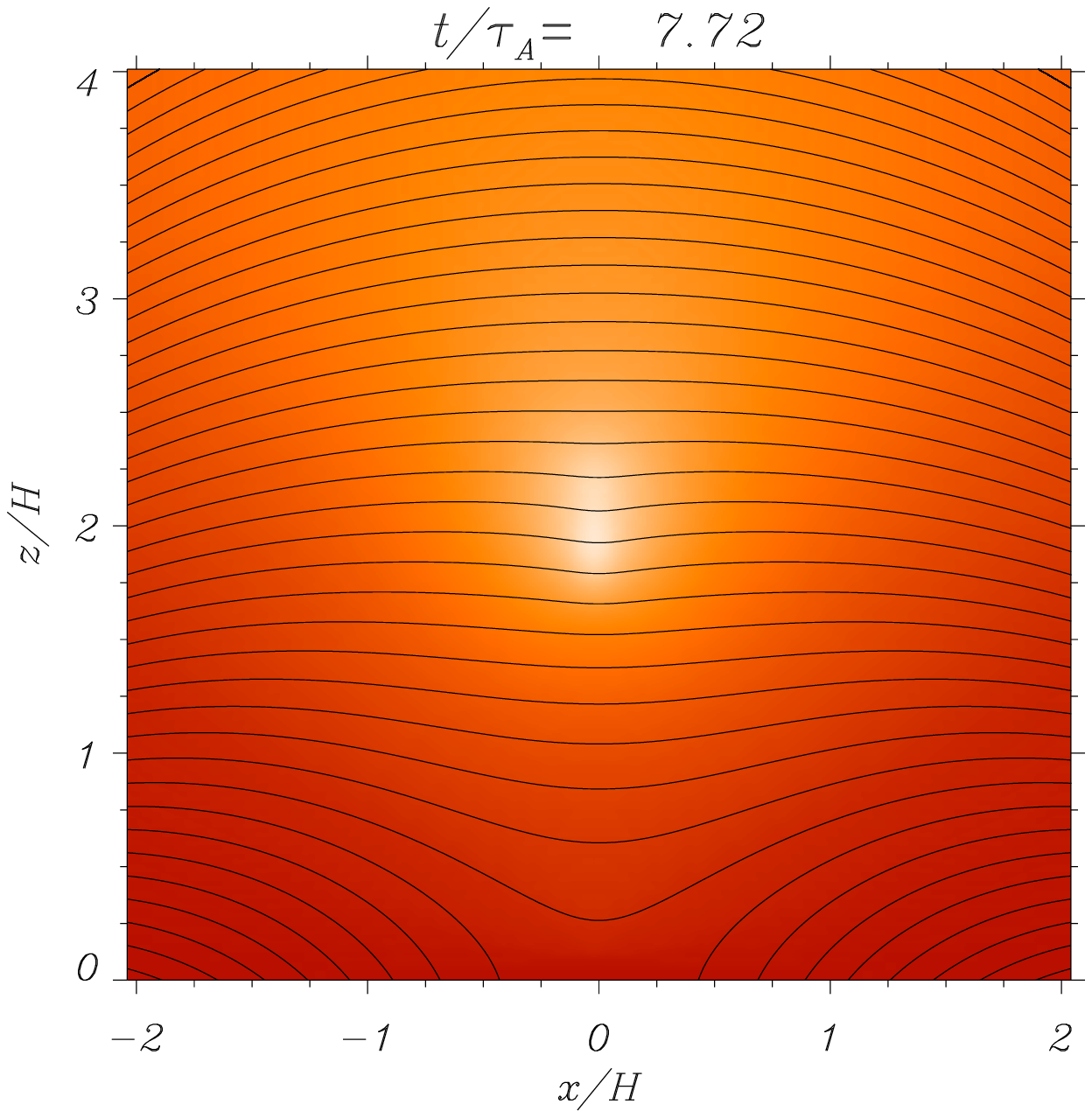}}
\center{\includegraphics[width=7cm]{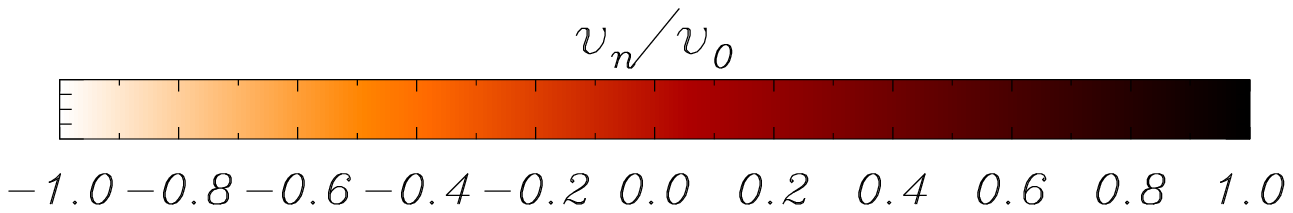}} \caption{\small Velocity
normal to the field lines, $v_n$, for a fast transverse excitation at two different times around the prominence
body located at $x=0$ and $z=2\,H$ (see Fig.~\ref{denstemp} for the density
distribution).}\label{velvn}
\end{figure}

A  particular prominence model is first selected. Using the linear code a
perturbation at the prominence body is introduced in the vertical direction at a
given instant ($t=0$). For simplicity the spatial dependence of the velocity
perturbation is the same as in Eq.~(\ref{source}). In Fig.~\ref{velvn} we find
the 2D distribution of the velocities at a given time of the evolution.  The
initial disturbance excites mainly fast MHD waves since the motion is dominated
by the components normal to the magnetic field lines, while the parallel
component is rather small. In a low plasma-$\beta$, normal/parallel motions are
typically related to fast/slow MHD waves. We have used the following relations to calculate
the velocity components \begin{eqnarray}
v_n=v_x\,\frac{B_z}{B}-v_z\,\frac{B_x}{B},\label{v_n}\\
v_\parallel=v_x\,\frac{B_x}{B}+v_z\,\frac{B_z}{B}.\label{v_par} \end{eqnarray}
\noindent The velocity field indicates that the prominence is oscillating as a
whole mostly in the vertical direction. The spatial distribution of $v_n$ has a maximum at the
location of the prominence core but it shows a tail extending in the
$z-$direction.

\begin{figure}[!ht] \center{\includegraphics[width=8cm]{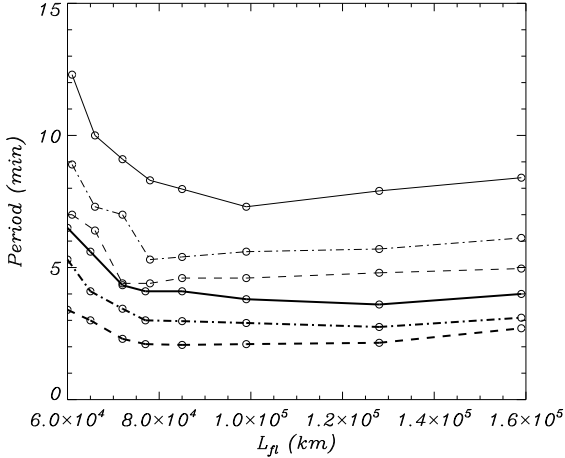}} 
\caption{\small Period of the fundamental vertical fast mode as a
function of the length of the field line crossing the center of the prominence.
Black curves correspond to a prominence with $M/L_y=5.4\times 10^{5}\,\rm kg\, km^{-3}$, 
dot-dashed curves to $M/L_y=2.7\times 10^{5}\,\rm kg\, km^{-3}$, and
$M/L_y=1.3\times 10^{5}\,\rm kg\, km^{-3}$ for the dashed curve. Thin curves are
associated to  $v_{\rm A0}=10\,c_{\rm s0}$, while thick curves correspond to
$v_{\rm A0}=20\,c_{\rm s0}$.}\label{p_L} \end{figure}

Hereafter we mostly focus on the period of oscillation. By performing a
periodogram of the signal at the center to the prominence we compute the
dominant period of oscillation. This has been repeated for different equilibrium
models. One of the parameters that has been changed is $L$ (half the length of
the arcade). In Fig.~\ref{p_L} the period of the vertical mode is plotted as a
function of $L_{\rm fl}$ (the relationship between $L_{\rm fl}$ and $L$ is found in
Fig.~\ref{p_L-Llf}). Other important parameters have been also changed, the
different line styles correspond to the three different total masses of the
prominence considered in this work, while thin lines correspond to
$v_{\rm A0}=10\,c_{\rm s0}$ and thick lines to $v_{\rm A0}=20\,c_{\rm s0}$, i.e., associated to
values of the magnetic field at the prominence core of 2 and 4 Gauss,
respectively.

Several conclusions can be extracted from Fig.~\ref{p_L}. We see that the period
of oscillation does not have a very strong dependence on $L_{\rm fl}$. For small
$L_{\rm fl}$ the periods are slightly longer than for large $L_{\rm fl}$, and in this
last regime the period attains an almost constant value. We also see that, as
expected, the period increases when the total mass of the prominence is raised.
For example, for large $L_{\rm fl}$ the differences in period between the lightest
prominence ($5 \,\rm min$) and the heaviest prominence ($8 \,\rm min$) is only
around $3\,\rm min$ for the case $v_{\rm A0}=10\,c_{\rm s0}$. Figure~\ref{p_L} also
indicates that the periods for $v_{\rm A0}=20\,c_{\rm s0}$ are shorter than those for
$v_{\rm A0}=10\,c_{\rm s0}$ by a factor which is around 2. These results suggest that
there might be a simple relationship of the period with the different
equilibrium parameters. To investigate this point we compare the results of our
simulations with simple analytical models whose explicit dependence of the
period on the equilibrium parameters is known. 

\begin{figure}[!ht] \center{\includegraphics[width=8cm]{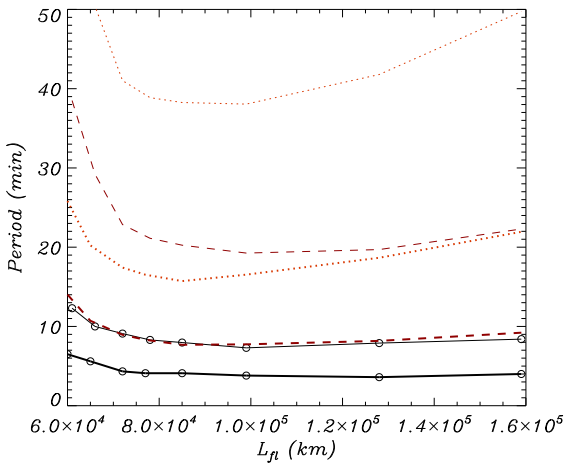}} 
\caption{\small Period of the fundamental vertical fast mode as a function of
the length of the field line crossing the center of the prominence. In this plot
$M/L_y=5.4\times 10^{5}\,\rm kg\, km^{-3}$.  Thin curves are associated to 
$v_{\rm A0}=10\,c_{\rm s0}$, while thick curves correspond to $v_{\rm A0}=20\,c_{\rm s0}$.
Dotted-orange curves represent the analytical approximation given by
Eq.~(\ref{omega_tt}), dashed-red curves correspond to the results for a finite
straight slab while continuous curves are the results of the full numerical
problem.}\label{p_Lcomp} \end{figure}

The first model we can use for the comparison is the infinite straight slab with
a transverse constant magnetic field. If the prominence body has a length $L_{\rm p}$,
and the total length of the field lines is $L_{\rm T}$, then for the situation $L_{\rm p}\ll
L_{\rm T}$ the period is \citep[see for
example][]{joarderroberts92,diazetal10,soleretal10} \begin{eqnarray} P\simeq \pi
\frac{1}{v} \sqrt{\left(L_{\rm T}-L_{\rm p}\right) L_{\rm p}},\label{omega_tt} \end{eqnarray}
where $v$ is the typical speed in the prominence body.  In the case of fast MHD
oscillations $v=v_{\rm Ap}$. We have calculated from the results of the simulations
the Alfv\'en speed at the center of the prominence body and have also estimated
$L_{\rm p}$ and $L_{\rm T}$ ($L_{\rm p} \approx w_x$ and $L_{\rm T} \approx
L_{\rm fl}$). Using these values the period is calculated from
Eq.~(\ref{omega_tt}). The results are shown in Fig.~\ref{p_Lcomp} with dotted
curves. These curves have to be compared with the results of the period inferred
from simulations shown in solid curves (thin curves correspond to the case 
$v_{\rm A0}=10\,c_{\rm s0}$ while thick curves represent the case $v_{\rm A0}=20\,c_{\rm s0}$).
There is a difference of more than a factor 4 between the simple analytical
periods and the ones inferred from the solution of the full problem. This
indicates that the simple slab model is not a good representation of our
prominence model. Among other things, in Eq.~(\ref{omega_tt}) an infinite
extension of the prominence in the vertical direction is assumed.

A way to improve the comparison is to evaluate the role of a finite height of
the slab on the period of oscillation. To do this we have to recall the results
of \cite{diazetal01} who studied the problem of fast MHD waves in a finite
two-dimensional slab representing a prominence thread. The magnetic field in
their model is constant and density changes abruptly between the corona and the
prominence. The model does not take into account the effect of gas pressure
($\beta=0$), and gravity and curvature of the magnetic field are neglected. The
authors derived a general dispersion relation based on an infinite system of
homogeneous algebraic equations. For comparison purposes we have used the method
of \cite{diazetal01} to calculate the eigenfrequencies of oscillation using the
equilibrium values derived from our simulations which include many effects that
are still missing in the model of these authors. The results of these
calculations are shown in Fig.~\ref{p_Lcomp} with dashed lines. We see a
significant decrease of the period in comparison with the unbounded slab (around
a factor 2). Notice that the behavior for large $L_{\rm fl}$ for the finite slab is
very similar to the profile of the numerically calculated periods for the full
case. Therefore, the finite slab model incorporates an effect that improves the
comparison with the full numerical case. Nevertheless, the differences with
respect to the numerical simulations remain significant. One of the assumptions
of the finite slab problem is that $\beta$ is zero, and this is not the case for
the full problem. A simple way to asses the effect of finite $\beta$ is to
compare the results for the two values of the Alfv\'en speed used in this work.
Figure~\ref{p_Lcomp} shows that the difference between the periods of the finite
slab model and the numerical results decreases as $\beta$ is decreased. For
$v_{\rm A0}=10\,c_{\rm s0}$ this ratio of the periods is typically 2.5 while for
$v_{\rm A0}=20\,c_{\rm s0}$ is around 2. 

The effect of curvature of the magnetic field can also play a role in the
discrepancy between the finite slab model and the full problem. In this regard,
it was shown by \cite{diazetal06,diaz06} that curved loops with cylindrical and
elliptical geometries have periods for the fundamental vertical fast mode that
are always smaller than that of the equivalent straight models. This is in
agreement with the behavior found in Fig.~\ref{p_Lcomp}. Although the results of
\cite{diazetal06,diaz06} are for a tube which is fully filled and here we
consider a tube that is only partially filled by the prominence core, these
theoretical works point out that curvature typically lowers the period in this
sort of 2D configurations. It is interesting to point out that in curved
configurations most of the modes have a leaky character, and the global fast mode
studied here is not an exception. More details about this issue are given in
Section~\ref{alfsec}.

So far the analytical modes used in our comparison assume that the equilibrium
magnitudes are piecewise constant along the magnetic field. This is certainly
not true in our numerical prominence models (see for example Fig.~\ref{va_cs_L})
and this might introduce also some changes on the period of oscillation.
Unfortunately the investigation of this effect is not straight forward since up
to know the analytical works in this direction have been mainly focused on
straight and fully filled cylinders representing coronal loops \citep[see for
example][]{andriesetal05a,andriesetal05,diazetal06a}. In any case, it is known
that, at least for the fundamental fast MHD mode, what really matters regarding
the period of oscillation is the value of the equilibrium magnitudes around the
center of the structure while the changes around the footpoints are less
important because of the line-tying boundary conditions.

Gravity has been ignored in the analysis of prominence oscillations in most of
the previous analytical works. Here we have a simple way to check the role of the
gravity force on fast MHD waves. Since the periods of oscillation are calculated
using the linearized set of equations we have compared these values of the
periods when the gravity term is included and when it is set to zero. The
differences are really small,  of the order of $2\%$ only, and this leads us
to conclude that the effect of gravity on fast MHD modes in prominences is very
small at least for the range of parameters considered in this work. 


Another question that we have addressed is whether the shape of the prominence
has a strong influence on the period of the vertical oscillation. We have
performed different experiments changing the prominence from vertical to
horizontal, but keeping the same mass and geometrical aspect ration. We have 
concentrated on models with the longest size of the field lines. The results of
the simulations clearly indicate that the period of the vertical mode is
essentially the same. Thus, the particular shape of the prominence seems not to
be very important regarding the period of vertical oscillations (at least in the
Cartesian geometry studied here). We have also carried out the same experiment
but using the model of \cite{diazetal01} and have arrived to the same conclusion.
This is an interesting result, for fast MHD waves what really matters is the
amount of mass of the prominence and not its particular geometrical shape. This
conclusion might not be true in cylindrical geometry and also for the
longitudinal harmonics in Cartesian geometry.

In summary, we deduce that the deviation of the actual period of vertical
oscillation from the prediction of Eq.~(\ref{omega_tt}) is mainly due to the
effects of the finite vertical extent of the prominence and of curvature.
Finally, it is important to mention that no hints of instability have been found
when vertical oscillations are excited in the system. In principle instability is
possible in this configuration since at the bottom of the prominence we have a
situation of a heavy plasma on top of a light plasma under the presence of
gravity \citep[see][for an application to prominence threads]{terretal12}.
However, magnetic tension is sufficient to counteract gravity making the
interface stable. According to the dispersion relation of the interface
\citep[see for example][]{chandra61} if perturbations are assumed to have a component along the
$y-$direction then Rayleigh-Taylor instabilities would be enhanced since the
gravitational term increases.

\subsection{Longitudinal waves}

\begin{figure}[!ht] \center{\includegraphics[width=7cm]{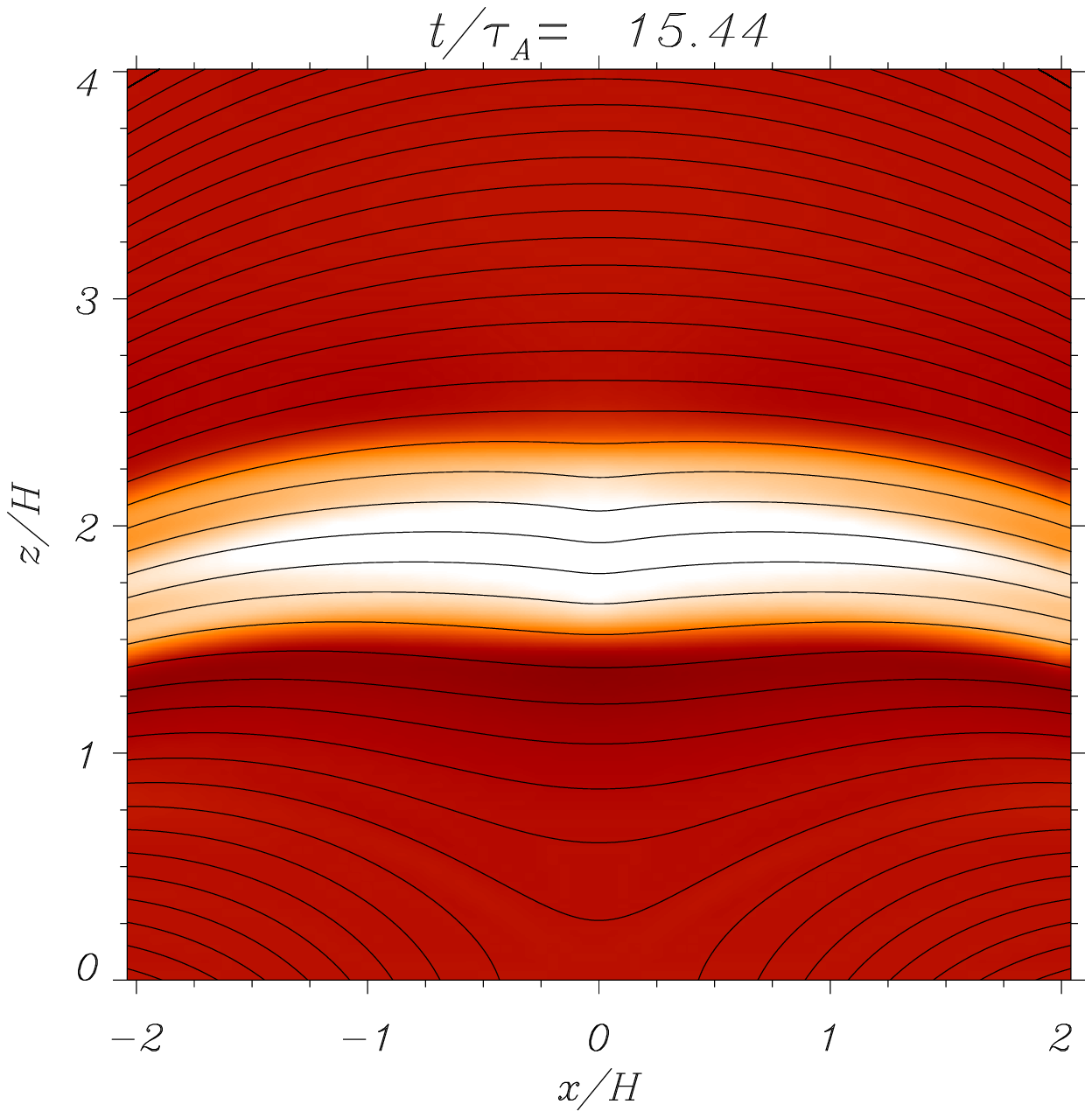}}
\center{\includegraphics[width=7cm]{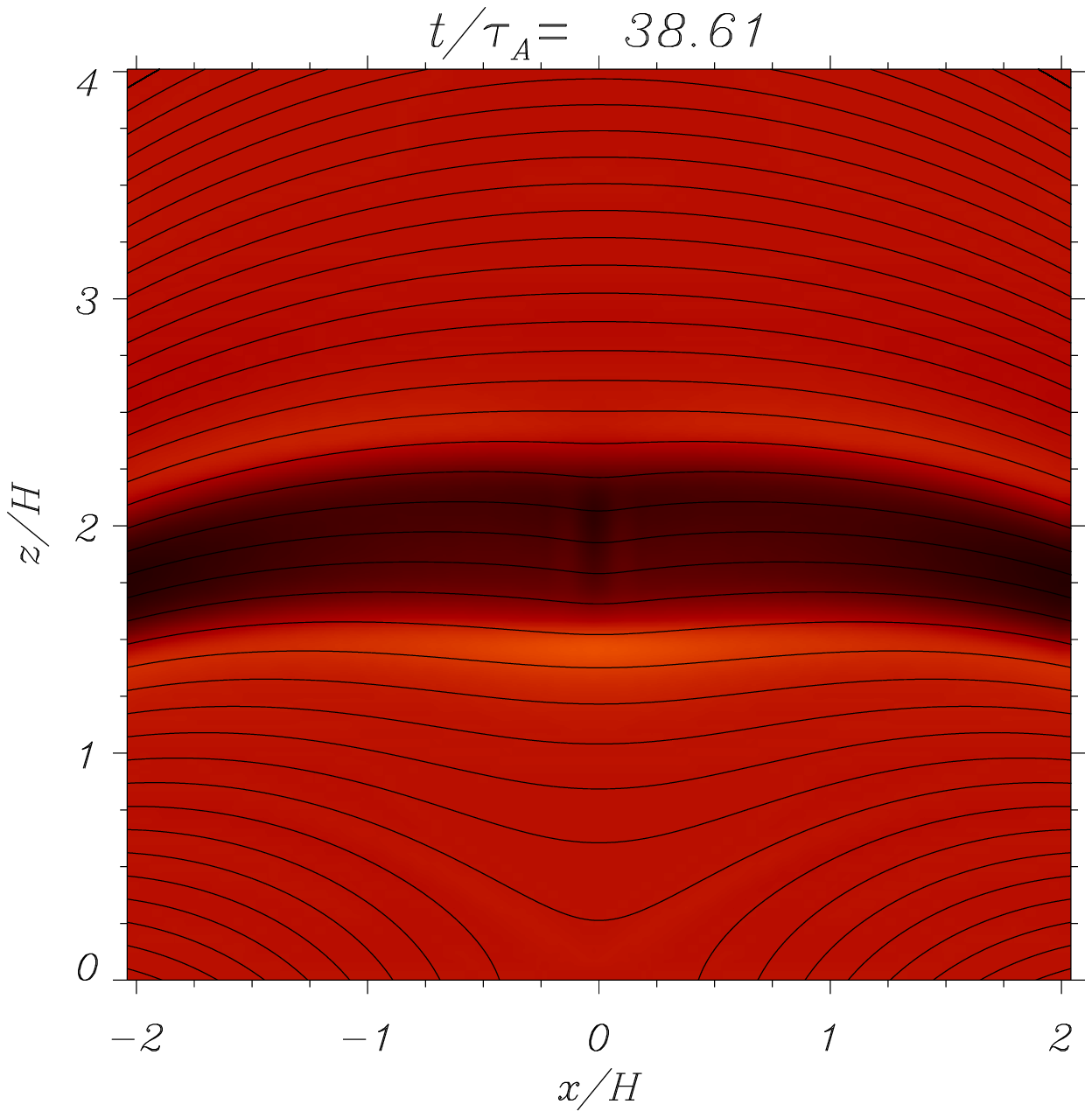}}
\center{\includegraphics[width=7cm]{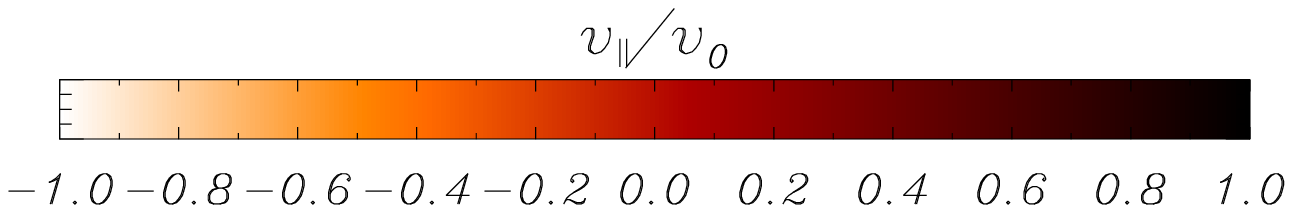}} \caption{\small Same as
in Fig.~\ref{velvn} but for the velocity component parallel to the magnetic
field lines, $v_\parallel$, for
 a slow mode excitation.}\label{velvpar}
\end{figure}

Slow magnetoacoustic-gravity  modes are investigated in this Section. In our
model, the spectrum of slow-gravity modes contains a continuum of frequencies plus
discrete modes. We first focus on discrete modes by introducing a horizontal
perturbation exciting the whole prominence body. Since the motions are basically
polarized along the magnetic field lines we call these motions longitudinal.
Snapshots at two different times of the velocity component parallel to the
magnetic field, calculated using Eq.~(\ref{v_par}), are found in
Fig.~\ref{velvpar}. The distribution of $v_\parallel$ shows a strong localization at the prominence
body  but also along the field lines that pass through the core and connect to
the photosphere where line-tying is imposed (not shown in Fig.~\ref{velvpar}).

\begin{figure}[!ht] \center{\includegraphics[width=8cm]{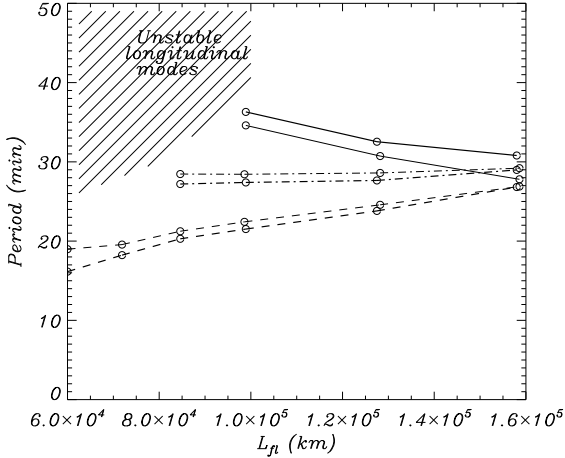}} 
\caption{\small Same as in Fig.~\ref{p_L} but for slow-gravity
modes.}\label{period_Ls} \end{figure}

\begin{figure}[!ht] \center{\includegraphics[width=8cm]{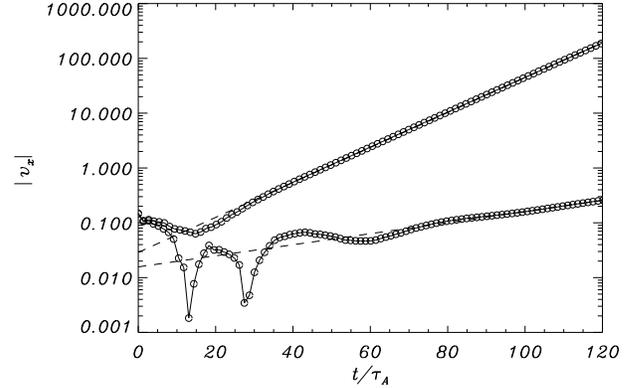}} 
\caption{\small Horizontal velocity at the center of the prominence as a function
of time for two different configurations. The upper curve corresponds to the case
$L=3\,H$, while for the lower curve $L=4\,H$. The vertical axis is in logarithmic
scale, meaning  that for large times the velocity grows exponentially. Dashed
curves are fits to the linear behavior of the curves and provide an estimation of
the growth-rate of the unstable modes ($\tau_{\rm g}=13\,\tau_{\rm A}$ for the
upper curve and $\tau_g=42\,\tau_{\rm A}$ for the lower
one).}\label{unstable_long} \end{figure}

The period of oscillation of discrete slow magnetoacoustic-gravity waves, or for
short, longitudinal modes, is plotted in Fig.~\ref{period_Ls} and is longer than
that of the transverse vertical modes. This is an expected result since we are
in a low $\beta$ regime and therefore slow modes have lower frequencies (longer
periods) than fast modes. Now the periods associated to  the models with
$v_{\rm A0}=20\,c_{\rm s0}$, are a bit longer than those for $v_{\rm A0}=10\,c_{\rm s0}$, but
the difference is not 2 because for low $\beta$ slow modes the characteristic
velocity is essentially the sound speed, which does not change much for the two
reference Alfv\'en velocities considered here. Note that the profile of the
curves is different depending on the total mass of the prominence. Light
prominences show an increase of the period with the length of the magnetic field
line, while heavier prominences display the opposite behavior, the period is
decreasing with $L_{\rm fl}$. Moreover, for the two heaviest prominences the system
becomes unstable to longitudinal oscillations (see  dashed area in
Fig.~\ref{period_Ls}) for values of $L_{\rm fl}$ smaller than a critical threshold.
The unstable modes are characterized by an exponential increase with time of the
amplitude of all the perturbed variables. Two examples of such behavior are
plotted in Fig.~\ref{unstable_long} for different lengths of the arcade. The
physics behind this instability is basically that, if the dip is not big enough,
the mass of the prominence is able to fall down by the effect of the gravity
force along the field lines connecting to the photosphere. Thus the parameter
$L_{\rm fl}$, and therefore $L$, plays a relevant  role regarding the stability of
the structure with respect to basically longitudinal motions, since it
indirectly determines the size of the dip in the model. Figure
\ref{fieldlines_L} provides a good example, the narrow arcade model is unstable
(it falls in the unstable region plotted in Fig.~\ref{p_L}), while the
prominence in the wider arcade is stable with respect to longitudinal motions.
We have not performed a detailed study about the growth rates since the
instability is simply linked to the fact that the mass falls toward the base of
the corona.

 Note also from Fig.~\ref{period_Ls} that heavy prominences have longer
periods than light prominences. This is in agreement with the results of 
\citet{zhangetal2013} who simulated the formation of a prominence and analyzed
the periods of the corresponding slow modes using a 1D model (see their Figs.~3
and 4). The fact that the period increases when the mass is increased can be
explained by the decrement of the sound speed in the prominence body. If gravity
terms are not very important then the motion of the prominence is governed by
pressure forces and the frequency of oscillation is basically proportional to the
internal sound speed. Therefore, heavy prominences, associated to low values of
the sound speed, have longer periods than light prominences with higher values of
the sound speed (if gas pressure is kept constant).

\begin{figure}[!ht] \center{\includegraphics[width=8cm]{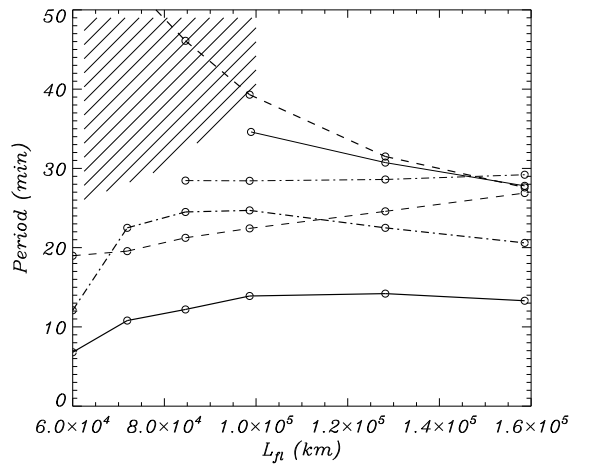}} 
\caption{\small Same as Fig.~\ref{period_Ls} but for the case
$v_{\rm A0}=10\,c_{\rm s0}$ only and including with thick curves the period calculated
analytically due to gravity ($P=2\pi\sqrt{R/g}$). }\label{period_Lsan} \end{figure}

Concerning slow magnetoacoustic-gravity modes \citet{lunakarpen12,lunaetal12}
claim that longitudinal oscillations are mostly driven by gravity and have a
period given by $P=2\pi\sqrt{R/g}$ where $R$ is the radius of curvature of the
dipped magnetic field. We have calculated the radius of curvature of different
prominence models and have computed the corresponding period using the previous
formula. The period from simulations together with the period due to gravity
only are displayed in Fig.~\ref{period_Lsan} for the case $v_{\rm A0}=10\,c_{\rm s0}$.
It is clear that the agreement between the two results is not good. The curves
associated to the analytical expression show a completely different behavior
with $L_{\rm fl}$. The analytical expression predicts that the heaviest prominence
should have a shorter period than that of the lightest one, while the results
from the simulations indicate the opposite dependence. These differences suggest
that the identification of longitudinal motions as purely due to gravity is not
appropriate at least in the present configuration. The reason of the discrepancy
is most likely due to the different assumptions made in the models. In
\citet{lunaetal12} the variation of the magnetic field along field lines is
neglected, their model is isothermal and more important  the radius of curvature
is constant. Since they consider  prominence threads the height of the plasma
column is quite short.


We turn to the modes of the continuum. \cite{goossensetal85} found expressions
for slow continua (and also Alfv\'en continua) in a general 2D magnetostatic
equilibrium with invariance in the $y-$direction (see their Eqs.~(59) and (60)).
These expressions were derived using a local orthogonal system of flux
coordinates, which for simplicity, are written here in terms of the distance
along the field lines, denoted by $s$. For slow modes, purely polarized along
the magnetic field lines for the situation $k_y=0$, we have
\begin{eqnarray}\label{slowcont} 
\frac{d^2\xi_s}{ds^2}+F(s)\frac{d\xi_s}{ds}+G(s,\omega) \xi_s=0,
\end{eqnarray}
where 
\begin{eqnarray}\label{Aslowcont} 
F(s)=\frac{2}{c_{\rm T}}\frac{d
c_{\rm T}}{ds}+\frac{d\ln{\rho}}{ds}-\frac{d\ln{B}}{ds},
\end{eqnarray}
\begin{eqnarray}\label{Bslowcont} 
G(s,\omega)=&-&\frac{1}{v_{\rm A}^2}g_s\left(\frac{d\ln{\rho}}{ds}-\frac{g_s}{c_{\rm s}^2}\right)+
\frac{1}{c_{\rm T}^2}\frac{d}{ds}\left(\frac{c_{\rm T}^2}{c_{\rm s}^2}\right) g_s\nonumber \\ &+&
\frac{1}{c_{\rm s}^2}\left(g_s\frac{d\ln{B}}{ds}+\frac{d g_s}{ds}\right)\nonumber \\
&-&\frac{d\ln{B}}{ds}\left(\frac{2}{c_{\rm T}}\frac{d
c_{\rm T}}{ds}+\frac{d\ln{\rho}}{ds}\right)
-\frac{d^2\ln{B}}{ds^2}\nonumber\\
&+&\frac{\omega^2}{c_{\rm T}^2}.
\end{eqnarray}
Here $c_{\rm T}$ is the tube or cusp speed, defined as $c_{\rm s}\,v_{\rm A}/\sqrt{c_{\rm s}^2+v_{\rm A}^2}$, $g_s$ is the projection of gravity along the
magnetic field line, and $\omega$ is the frequency that we want to calculate
(perturbations of the form $e^{i \omega t}$ have been assumed). Note that the different terms involve
derivatives of density, gravity, and magnetic field along the field lines.

Equation (\ref{slowcont}) must be complemented with appropriate boundary
conditions, which are line-tying at the two ends of the magnetic field lines.
Solving this equation and determining  the value of the eigenfrequency $\omega$
is not straight forward since our configuration does not have a simple geometry.
The numerical procedure is the following, a particular footpoint is selected and
the corresponding field line coordinates are calculated using the equilibrium
that has been determined numerically. Once the coordinates of the field line are
known, all the necessary variables are projected along this magnetic field line
as a function of the distance $s$ along the field line. The last step is to
numerically solve Eq.~(\ref{slowcont}) using the interpolated values.

\begin{figure}[!ht] \center{\includegraphics[width=8cm]{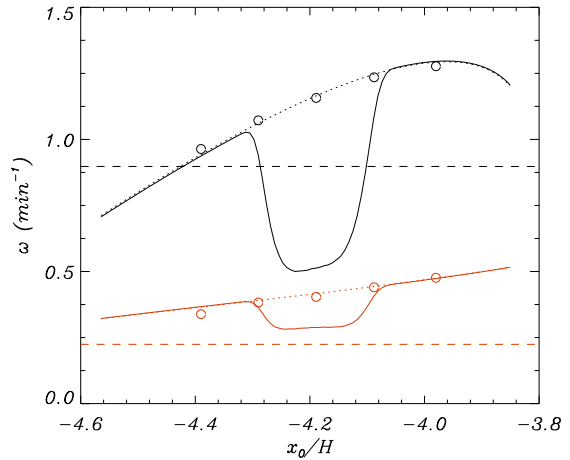}} 
\caption{\small Alfv\'en frequency (black curve) and slow frequency (red curve)
of the continuum, calculated using Eqs.~(\ref{alfvcont}) and (\ref{slowcont})
respectively, as a function of the footpoint position in a given range. Dots
correspond to the spectrum calculated without the heavy prominence, while circles
represent the values inferred from the time-dependent simulations also without
the dense prominence. Dashed lines correspond to the frequencies of the
fundamental vertical (high frequency) and longitudinal modes (small
frequency) inferred from the time-dependent simulations. In this plot $L=5H$,
$v_{\rm A0}=10\,c_{\rm s0}$, and $M/L_y=2.7\times 10^{5}\,\rm kg\,
km^{-3}$.}\label{alfslowfreq} \end{figure}

In Fig.~\ref{alfslowfreq} the computed frequency for slow MHD modes belonging to
the continuum, $\omega_{\rm C}$, is represented as a function of the footpoint
position in a fixed interval. The curve in dots in Fig.~\ref{alfslowfreq}
corresponds to the eigenmode calculations without a dense prominence and has
been included for comparison purposes. These eigenmode computations show a good
agreement with the results of the time-dependent problem, represented in
Fig.~\ref{alfslowfreq} with circles, when the initial excitation is localized on
different magnetic surfaces, producing an efficient slow mode excitation. From
the agreement between the time-dependent results and the eigenmode results we
conclude that our solutions to Eq.~(\ref{slowcont}) are correct. The inclusion
in the arcade of a dense prominence produces the existence of a minimum in the
cusp speed. It is interesting to note that the discrete slow modes have
frequencies that are below this minimum. This is surprising since slow modes in 
homogeneous slabs \citep[][]{edrob82} have frequencies between internal cusp
speed and the internal sound speed (under coronal conditions though). This
phenomena needs to be studied in more detail in future studies.

\subsection{Alfv\'en waves}\label{alfsec}

In this work we are not interested in the analysis of motions in the ignorable
direction which are related to the excitation of pure Alfv\'en waves. These modes
lack of a global character and cannot produce coherent global motions of the
prominence. However, we can
derive useful information from the corresponding eigenmode calculations. 
If $k_y=0$, we have the
following second order differential equation \citep[see][]{goossensetal85,
oliveretal93} for the modes of the Alfv\'en continuum
\begin{eqnarray}\label{alfvcont}  \frac{d^2\xi_y}{ds^2}+\frac{d\ln{B}}{ds}
\frac{d\xi_y}{ds}+\frac{\omega^2}{v_{\rm A}^2} \xi_y=0, \end{eqnarray} where $\omega$ is the
eigenfrequency. Note that there is no dependence with gravity in this equation and that
there is a term which accounts for the variation of the strength of the magnetic field
along field lines. The computation of the Alfv\'en spectrum is useful in order to
understand the damping of global transverse modes. From the numerical point of view we proceed
in a similar way as for the slow modes belonging to the continuum.

In Fig.~\ref{alfslowfreq} the computed frequency for Alfv\'en modes, $\omega_{\rm
A}$, is represented as a function of the footpoint position. The Alfv\'en
frequency is larger than that of the slow or cusp frequency since we are in a
situation where magnetic pressure dominates over gas pressure. The agreement
between eigenmode calculations and the results of the time-dependent problem,
using a localized perturbation on magnetic surfaces in $v_y$, is evident (compare
circles with the dotted curve). Again the depression in frequency around the
footpoint located at $x_0=-4.2 H$ is due to the presence of the heavy prominence.
The frequency of the discrete fundamental vertical mode determined from the
linear time-dependent problem, $\omega_{\rm f}$, is also plotted in 
Fig.~\ref{alfslowfreq} as a horizontal dashed line. We see that depending on the
footpoint position the frequency of the global vertical mode is above or below
the local Alfv\'en frequency. If the frequency is above the Alfv\'en frequency it
means that the eigenfunction has an oscillatory behavior, while if it is below
its behavior is evanescent \citep[see for
example][]{bradarb05,bradetal06,veretal06,rialetal13}. The fact that $\omega_{\rm
f}>\omega_{\rm A}$ for $x_0<-4.3 H$, which corresponds to magnetic field lines
with apexes situated at higher heights as $x_0$ is decreased, means that the
eigenfunction is not confined and it is oscillatory. This is an indication that
the mode is unable to trap all the energy and has a leaky character, being its
energy radiated away from the prominence body. The leaky character of the mode
produces an attenuation of the amplitude with time, and this feature is already
present in Fig.~\ref{graph7}. It is known that the inclusion of perpendicular
wavenumber in the $y-$direction might significantly reduce this leakage. In
addition, we have to bear in mind that the plasma-$\beta$ changes significantly
in the computational domain (see Fig.~\ref{plasmabeta}) and there are regions
where the sound speed is equal to the Alfv\'en speed ($\beta$ close to one).
Under such conditions the character of the modes can change due to mode
conversion and contribute to the damping of the global modes. As the damping of
oscillations is not the main topic of this paper we have not investigated further
the mechanism of mode conversion.

From the numerical perspective both slow and Alfv\'en modes show a strong
attenuation with time which is produced by numerical dissipation. As the modes of
the continuum are localized on magnetic flux surfaces and a Cartesian grid is
used in the simulations, it is difficult to capture the correct spatial structure
of the modes but nevertheless we still get periods, as Fig.~\ref{alfslowfreq}
indicates, which are quite reliable. A possible way to improve in this aspect
is by using flux coordinates in the simulations. The reader is referred to
\citet{rialetal13} for an example of such kind of simulations in a curved
magnetic field.

\section{Discussion and conclusions}

A numerical method to obtain MHS solutions with some specific features that
mimic real prominences has been presented. The method is based on the injection
of mass through the continuity equation and on the relaxation of the system. The
initial equilibrium is given an increase in gravitational energy by the increase
of the density at a given location. Part of this energy is converted into an
increment of the magnetic energy and internal energy but also in an increase of
kinetic energy. We have found that our numerical scheme is suitable to study the
evolution of the system  towards a situation that is close to a stationary
state. The treatment of boundary conditions, based on a decomposition in
characteristic fields, is the key part of the relaxation method since the energy
excess is allowed to leave the system through the boundaries. By the time we
were preparing this manuscript we were aware of the work of 
\citet{hilliervan13}  who have also used a relaxation method in the context of
prominences. However, their study is based on  flux rope structures and their
relaxation technique uses over dissipation to achieve a stationary state. Thus,
the method used in our work, based on the decomposition in characteristics, is
conceptually different.

Using the relaxation method we have built a set of new prominence equilibrium
models in 2D. These models include the connection of the prominence body with
the photosphere, and contain a cool core that matches the internal temperature
with the coronal temperature through a PCTR. The size of the prominence core and
PCTR are determined by the choice of the spatial distribution of the mass source
term. The gravity force is included in the numerically generated models.

It is clear that the method devised here to find MHS solutions should be
extended to three-dimensional geometries. In particular, a future application
could be to use 3D magnetic field extrapolations of a real prominence to test
the global support of the structure by injecting mass in the magnetic
configuration and studying the time evolution. From the comparison of the model
with observations, some conclusions about the magnetic field extrapolations
could be extracted.

The second part of the work is about the analysis of MHD waves in the numerically
generated models. The properties of the different types of MHD modes have been
studied. Since we have restricted to oscillations without perpendicular
propagation in the $y-$direction ($k_y=0$) the different types of waves are
easily identified because they are uncoupled. We have found that vertical
oscillations, associated mainly to fast MHD waves, are always stable, at least
for the equilibrium parameters considered in this work, when there is no
perpendicular propagation. The obtained periods are typically in the 4-10 min
range. On the contrary, longitudinal oscillations, related to slow
magnetoacoustic-gravity waves have longer periods periods which are in the range
28-40 min. These waves are strongly affected by gravity and can become unstable
when short magnetic arcades are considered because they are unable to have
significant dips which are the key to have stable magnetic configurations. 

The two different groups of periods found in this work and related to fast and
slow MHD waves are below the values of the reported periods from observations.
For example, we can compare with the periods found in Table 1 of
\citet{tripatetal09} that are associated, in most of the cases, to motions of the
whole structure. In that table we can distinguish that periods of vertical
motions are typically in the range 15-29 min, while longitudinal motions (along
the filament axis) are in the range 50-160 min. We think that more realistic
configurations are necessary to improve the comparison and 3D models are the key
point to achieve this. In this regard, the information inferred from the
properties of MHD oscillations in prominence structures like the ones studied in
this work my lead in the near future to the application of prominence seismology
\citep[see for example][]{arreguietal12}.

The periods of oscillation in our complex configurations have been compared with
the periods predicted by simple models that miss many physical effects. Regarding
fast MHD waves we have found that, for example, the results of the infinite slab
only provide an estimation of the order of magnitude of the period of
oscillation. The effects of considering  a prominence with a finite height 
improves the comparison. Gravity seems to be not important for vertical motions
of the prominence, and interestingly, the geometrical shape of the prominence is
not relevant. On the contrary, slow magnetoacoustic-gravity waves have a
completely different behavior. They are strongly affected by the gravity force
and can even become unstable. The existence of a continuous spectrum of slow MHD
waves complicates the interpretation of the periods of oscillation since in
general there is a joint excitation of both discrete and continuum modes.

We expect that the inclusion of perpendicular propagation in the model can
significantly change some of the properties of the MHD waves. First of all, it
will produce the resonant damping of the global vertical mode with the modes of
the Alfv\'en continuum. Second, it can have  a strong effect on vertical modes
since they may become Rayleigh-Taylor unstable \citep[see][for recent results
about 3D modeling]{hillieretal11,hillieretal12}. Nevertheless, the inclusion of
shear and twist, ignored in the present work, might have an stabilizing effect.
As far as we know, MHD stability analysis in theoretical prominence models of
this type are very scarce. Thus, the problem of instabilities due to gravity
together with twist or shear should be carefully examined in future studies using
three-dimensional models. This is an interesting problem since the dynamics of
the instability could be exhaustively investigated and a more realistic
comparison with the observations could be attempted. 

\acknowledgements J.T. acknowledges support from the Spanish Ministerio de
Educaci\'on y Ciencia through a Ram\'on y Cajal grant. All the authors
acknowledge the funding provided under the project AYA2011-22846 by the Spanish
MICINN and FEDER Funds.  A.J.D. acknowledges the financial support by the Spanish
Ministry of Science through project AYA2010-18029. The financial support from
CAIB through the ``Grups Competitius'' scheme is also acknowledged. The authors
also thank M. Luna, G. Verth and A. W. Hood for their comments and suggestions
that helped to improve the paper. The comments of the anonymous referee are also
acknowledged.


\end{document}